\documentclass{ar-1col}
\usepackage{url}
\usepackage{xspace}
\usepackage{subcaption}

\jname{Xxxx. Xxx. Xxx. Xxx.}
\jvol{AA}
\jyear{YYYY}
\doi{10.1146/((please add article doi))}

\newcommand{\ie}{{\it i.e.}}
\newcommand{\eg}{{\it e.g.}}
\newcommand{\beq}{\begin{equation}}
\newcommand{\eeq}{\end{equation}}
\newcommand{\sqrts}{\ensuremath{\sqrt{s}}\xspace}
\newcommand{\sqrtsnn}{\ensuremath{\sqrt{s_{_{\mathrm{NN}}}}}\xspace}
\newcommand{\RAA}{\ensuremath{R_{\mathrm{AA}}}\xspace}

\newcommand{\pt}{\ensuremath{p_T}\xspace}
\newcommand{\lsim}{\raisebox{-4pt}{$\,\stackrel{\textstyle <}{\sim}\,$}}
\newcommand{\gsim}{\raisebox{-4pt}{$\,\stackrel{\textstyle >}{\sim}\,$}}
\newcommand{\Tpc}{$T_{\rm pc}$}
\newcommand{\Ds}{${\cal D}_s$}
\newcommand{\qhat}{\hat{q}}

\begin{document}

\markboth{Dong et al.}{Open Heavy Flavor}

\title{Open Heavy-Flavor Production in Heavy-Ion Collisions}

\author{Xin Dong,$^1$ Yen-Jie Lee,$^2$ and Ralf Rapp$^3$
\affil{$^1$Nuclear Science Division, Lawrence Berkeley National Laboratory, Berkeley, CA 94720, USA; email: xdong@lbl.gov}
\affil{$^2$Laboratory for Nuclear Science, Massachusetts Institute of Technology, Cambridge, MA 02142, USA; email: yenjie@mit.edu}
\affil{$^3$Department of Physics \& Astronomy and Cyclotron Institute, Texas A\&M University, College Station, TX 77843-3366, USA; email: rapp@comp.tamu.edu}}

\begin{abstract}
The ultra-relativistic heavy-ion programs at the Relativistic Heavy Ion Collider and the Large Hadron Collider have evolved into a phase of quantitative studies of Quantum Chromodynamics at very high temperatures. The charm and bottom hadron production offer unique insights into the remarkable transport properties and the microscopic structure of the Quark-Gluon Plasma (QGP) created in these collisions. Heavy quarks, due to their large masses, undergo Brownian motion at low momentum, provide a window on hadronization mechanisms at intermediate momenta, and are expected to merge into a radiative-energy loss regime at high momentum. We review recent experimental and theoretical achievements on measuring a variety of heavy-flavor observables, characterizing the different regimes in momentum, extracting pertinent transport coefficients and deducing implications for the ``inner workings" of the QGP medium.
\end{abstract}

\maketitle

\tableofcontents

\section{Introduction}
\label{sec:intro}
The relativistic heavy-ion program aims at studying Quantum Chromo-Dynamics (QCD) at finite temperature ($T$) and baryon density ($\rho_B$). Numerical computations of the QCD partition function on a discretized space-time lattice, referred to as lattice-QCD (lQCD), show that at high temperature and vanishing baryon chemical potential ($\mu_B$) a transition occurs from hadronic matter, where quarks and gluons are confined and the chiral symmetry of the QCD Lagrangian is spontaneously broken, to a deconfined and chirally symmetric Quark-Gluon Plasma (QGP). For a realistic set-up with two light- and one strange-quark flavor, the pseudo-critical transition temperature associated with the restoration of chiral symmetry is by now rather well established at around \Tpc=155-160\,MeV~\cite{Borsanyi:2010bp,Bhattacharya:2014ara,Bazavov:2014pvz,Ding:2015ona}. Over the past two decades, experimental results from the Relativistic Heavy Ion Collider (RHIC) at Brookhaven National Laboratory and the Large Hadron Collider (LHC) at CERN have collected remarkable evidence of the formation of the QGP and its properties, see, \eg, recent reviews in Refs.~\cite{Shuryak:2014zxa,Braun-Munzinger:2015hba,Busza:2018rrf}.

\begin{marginnote}[]
\entry{\Tpc}{pseudo-critical temperature for QGP formation, defined, \eg, as the inflection point of the chiral quark-antiquark condensate.}
\end{marginnote}

Based on the asymptotic freedom of QCD, \ie, the decrease of the strong coupling constant, $\alpha_s(Q^2)$, with momentum transfer ($Q^2$), one anticipates the QGP to be a weakly coupled plasma at high temperature where deconfined quarks and gluons can travel rather freely over large distances. However, observations from ultra-relativistic heavy-ion collisions (URHICs) conducted at RHIC and the LHC demonstrated that the bulk medium created in these reactions is strongly coupled and highly opaque. On the one hand, viscous hydrodynamic models can give a good description of the transverse-momentum ($p_{T}$, with respect to the ion beam axis) spectra of light hadrons ($\pi$, $K$, $p$) and their distribution in azimuthal angle for low $p_T\lsim$\,2--3\,GeV/$c$. This indicates a rapid local thermalization of the initially produced medium with a subsequent development of collective flow driven by the pressure gradients in the system, requiring a very small specific viscosity, \ie, ratio of shear viscosity over entropy density, $\eta/s$. Quantitative fits to the light-hadron $p_T$ spectra and their azimuthal distributions have extracted $\eta/s$ values of around 0.1--0.3~\cite{Heinz:2013th,Gale:2012rq}, close to the conjectured quantum lower bound of 1/4$\pi$~\cite{Policastro:2001yc}, but its temperature dependence remains under debate~\cite{Paatelainen:2013eea}. On the other hand, the observed suppression of high-$p_T$ hadron production in nucleus-nucleus (A+A) relative to proton-proton ($p$+$p$) collisions has been successfully described by the energy loss of partons traversing the QGP using perturbative-QCD (pQCD) calculations; estimates of the associated jet transport parameter, $\qhat=\langle \Delta p_T^2\rangle/\lambda$, characterizing the average transverse-momentum broadening per unit path length of the parton, yield $\qhat/T^3 \simeq$ 3--6~\cite{Burke:2013yra} at a parton energy scale of 10\,GeV. The reconciliation of the such extracted strongly and weakly coupled features of QCD matter at large and small wavelengths, respectively, to robustly characterize the transition from one regime to the other, and unraveling the underlying microscopic mechanisms, have become central objectives of the relativistic heavy-ion collision program, in a concerted effort of experiment and theory. 

For light- and strange-flavor hadrons, it is rather challenging to develop and apply transport approaches that describe their production over the full kinematic $p_T$ region~\cite{Lin:2004en,Xu:2004mz,Cassing:2008sv}. The small masses of light partons facilitate conversions among them, and their strong coupling at small momentum transfers suggests that the quasi-particle approximation inherent to the semi-classical Boltzmann equation is no longer valid. For example, a thermalization time of $\tau_0$=1~fm/$c$ (assumed to be even smaller in most hydrodynamic models) implies a scattering rate of $\sim$3/fm, translating into a collisional width of 600\,MeV, quite comparable to the typical thermal masses of partons in the QGP. Furthermore, once they thermalize (as the low-$p_T$ light-hadron spectra suggest), the memory of the thermalization process is lost. 

Heavy quarks are considered ``heavy" for two reasons: first, in the particle physics context, their mass is larger than the typical nonperturbative scale of QCD, $m_Q \gg \Lambda_{\rm QCD}$, which enables the evaluation of their production cross sections within pQCD~\cite{FONLL}. Second, in the context of QCD matter formed in URHICs, their mass is larger than the typical temperature reached in the ambient medium, $m_Q \gg T_{\rm QGP}$; this implies that heavy-quark (HQ) production is essentially restricted to the initial hard scatterings (with a rather short formation time of $\sim1/2m_Q\sim0.1$\,fm/$c$), and their thermalization time becomes comparable to (or even larger than) the fireball lifetime~\cite{Svetitsky:1987gq,Rapp:2009my}. Therefore, heavy-flavor (HF) observables offer unique and comprehensive insights into the nature of the QGP medium and its hadronization, as further detailed below.




A key feature of HF probes in URHICs is their comprehensive coverage in transverse momentum, enabling systematic investigations of how the prevalent processes vary in different regions of $p_T$. At low $p_T$, the large mass of heavy quarks enables to treat their propagation in the QGP as ``Brownian motion",  with relatively small momentum transfers, of the order of the temperature, $q^2 \sim T^2$~\cite{Svetitsky:1987gq,Rapp:2009my}. Since energy transfer is parametrically suppressed by $T/m_Q$, elastic interactions dominate; the HQ motion can be reliably described by a stochastic Langevin process characterized by a (long-wavelength) transport parameter -- the HQ spatial diffusion coefficient, \Ds (See Chapter 2 for the definition). It has been predicted that charm quarks, despite their much larger mass compared to light quarks, can acquire significant collective radial and anisotropic flow when diffusion through the QGP~\cite{vanHees:2004gq,Moore:2004tg,vanHees:2005wb,Gossiaux:2008jv,Gossiaux:2009mk,He:2011qa,Cao:2011et,Alberico:2013bza,Berrehrah:2014kba,Song:2015sfa,Nahrgang:2016lst,Scardina:2017ipo,Plumari:2017ntm}. Experimentally, this was first found in pioneering measurements of semi-leptonic electron decay spectra at the Relativistic Heavy Ion Collider (RHIC)~\cite{Adler:2005xv,Adare:2006nq,Abelev:2006db}, and later confirmed and quantified at both RHIC and the Large Hadron Collider (LHC). In combination with the factor $\sim$3 heavier bottom quarks, low-$p_T$ HF diffusion therefore provides an excellent window on QCD matter in the nonperturbative regime and in this way enables insights into the ``inner workings" of its near-ideal liquid properties. In particular, the dimensionless scaled HF diffusion coefficient, $2\pi T$\Ds, is believed to carry universal information about the QGP transport properties, similar to $\eta/s$ or the electromagnetic conductivity, $\sigma_{\rm EM}/T$.

\begin{marginnote}[]
\entry{${\cal D}_s$}{spatial heavy-flavor diffusion coefficient, to be distinguished from the notation for charm-strange mesons, $D_s$.}
\end{marginnote}

At sufficiently high $p_T$, the mass effect ceases and HF observables should degenerate with those for light flavors. With decreasing $p_T$, however, a ``dead-cone" is expected to open up, \ie, a suppression of small-angle gluon radiation~\cite{Dokshitzer:2001zm}, suggesting an energy loss hierarchy of the type $\Delta E_b<\Delta E_c<\Delta E_q<\Delta E_g$, leading to less suppression of HF hadrons (and their decay products) compared to light-flavor hadrons~\cite{Buzzatti:2011vt,Andronic:2015wma}. 
The first measurements of HF decay electrons showed, however, that their $R_{\rm AA}$ is comparable to that of light flavor hadrons out to $p_T$'s of near 10\,GeV/$c$~\cite{Adler:2005xv,Adare:2006nq,Abelev:2006db} (note that the decay electrons typically carry less $p_T$ than the parent hadron). This indicated the important role of elastic energy loss in the QGP medium, which is not easily discernible using light-flavor probes. However, the expected energy loss hierarchy is likely affected by differences in initial parton spectra and hadronization processes~\cite{Djordjevic:2013pba}, and will cease at high $p_T$. Open HF probes offer a unique opportunity to systematically investigate the interplay of radiative and collisional energy loss mechanisms over a broad momentum region and identify the transition between the two.

\begin{marginnote}[]
\entry{Rapidity}{$y\equiv 0.5\ln((E+p_z)/(E-p_z))$, characterizes the longitudinal velocity distribution of particles of energy $E$ and 
momentum component $p_z$.}
\end{marginnote}
To quantify the modifications hadron spectra in A+A relative to $p$+$p$ collisions, two observables are widely used. The first one is the nuclear modification factor, 
\beq
R_{\rm AA} (p_T)= \frac{1}{N_{\rm coll}} \frac{d^2N_{\rm AA}/(dp_{\rm T}dy) } 
{ d^2N_{pp}/(dp_{\rm T}dy)} \ ,
\label{eq:raa}
\eeq
which is primarily used for particles which are produced at high momentum transfer in the primordial nucleon-nucleon collisions (such as high-\pt hadrons and heavy quarks). The production of these particles in A+A collisions is expected to scale with the number of primordial binary nucleon-nucleon collisions, $N_{\rm coll}$ (usually expressed as the product of the nuclear thickness function $T_{\rm AA}$, which characterizes the overlap density of nucleon-nucleon collisions in the incoming nuclei at given impact parameter, and the total inelastic cross section in $p$+$p$ collisions,  $N_{\rm coll}=T_{\rm AA} \sigma^{\rm inel}_{pp}$). Thus, in the absence of any medium effects, the \RAA
will be one, while energy energy loss phenomena as described above will lead to a suppression below one. The second class of common observables are anisotropic ``flow" parameters, $v_n$, defined as Fourier coefficients of the  azimuthal-angle distributions of final-state hadron spectra with respect to the ``event plane" of a given collision, 
\beq
\frac{d^3N}{p_{\rm T}dp_{\rm T}dyd\phi} = \frac{d^2N}{2\pi p_{\rm T}dp_{\rm T}dy} \bigg[ 1 + \sum_{n=1}^{\infty} 2v_n\cos\big(n(\phi-\Psi_{\rm EP})\big) \bigg] \ .
\label{eq:v2}
\eeq
The event plane is defined as the plane transverse to the direction of the incoming ion beams (usually defined as $z$-axis). It is azimuthally oriented in such a way that the impact parameter of the colliding nuclei is aligned with the event plane angle, $\Psi_{\rm EP}$ (\ie, for $\Psi_{\rm EP}$=0, $\phi$ is simply the angle of a hadron's momentum relative to the positive $x$-axis). The $n$-th harmonic flow coefficient, $v_n$, is then given by $v_n = \langle\cos(n(\phi-\Psi_{\rm EP}))\rangle$. Various experimental techniques are being applied to extract the $v_n$, \eg, by a direct event plane reconstruction or via scalar-product methods which avoid possible biases in the determination of the event plane.  
The most important (and largest) harmonic is the ``elliptic flow" coefficient, $v_2$. For a non-central A+A collision, which is characterized by an ``almond-shaped" spatial overlap region (with the short axis in $x$-direction), two basic sources of elliptic flow can be distinguished. The first one is related to pressure-gradients in a quickly thermalizing medium, which are larger along the short axis and thus generate a larger flow in the collective fireball expansion; this effect, generally leading to a positive $v_2$, will be prevalent at low $p_T$, in the bulk production of light hadrons, and has been a pivotal observable in extracting the medium's viscosities. The other one is related to path length differences experienced by high-$p_T$ particles traversing the medium; the shorter path length in $x$-direction generally leads to less suppression than for paths along the $y$-direction, thus also generating a positive $v_2$; high-$p_T$ $v_2$ is therefore a complementary observable in characterizing energy loss mechanisms, in particular their non-trivial path length dependence caused, \eg, by interference effects in gluon radiation.

The present review focuses on open HF hadrons, \ie, mesons and baryons that carry one charm or bottom quark or anti-quark. More than 95\% of the produced heavy quarks in heavy-ion collisions at RHIC and the LHC hadronize into these hadrons. The remaining fraction goes into either quarkonium states ($Q\overline{Q}$) or multiple-HF hadrons (\eg, $\Xi_{cc}$ baryons or $B_c$ mesons) in which the in-medium interactions between two heavy quarks can be probed to provide complementary information on the QGP medium; this sector will not be explicitly covered in this review. 
The discussion in the present article is organized into HQ diffusion in the low-momentum regime (Sec.~\ref{sec:diff}) with emphasis on elliptic flow as a key gauge of the pertinent transport coefficient, HQ hadronization (Sec.~\ref{sec:hadron}) highlighting the role of different species of charm and bottom hadrons, energy loss at high momentum (Sec.~\ref{sec:eloss}) addressing questions of the transition from collisional to radiative mechanisms and evidence of a mass hierarchy, implications of the HF results for the structure and hadronization of the QGP (Sec.~\ref{sec:qgp}), and the role of heavy flavor in small collision systems (Sec.~\ref{sec:pA}). The conclusions (Sec.~\ref{sec:concl}) consist of a compact set of summary points and future issues.


\begin{textbox}[h]\section{Heavy-quark production in $p$+$p$ collisions}
Measurements of open HF hadron production in proton-proton ($p$+$p$) collisions aim at testing perturbative-QCD (pQCD)~\cite{FONLL} and effective-field theory approaches~\cite{Kang:2016gfp}, and are crucial for establishing a baseline reference for measurements in heavy-ion collisions (recall Eq.~(\ref{eq:raa})). Inclusive charm- and bottom-hadron production cross sections measured at RHIC and the LHC are now providing robust benchmarks to assess cold- and hot-QCD matter effects in $p/d$+A and A+A collisions. Comparisons to Fixed-Order-Next-to-Leading-Logarithm (FONLL) pQCD calculations show that nearly all experimental data for both charm- and bottom-hadrons are at the upper end of the theoretical uncertainty bands, where the latter shrink appreciably at high $p_{\rm T}$~\cite{Adamczyk:2012af,Acosta:2003ax,Acharya:2017jgo,Sirunyan:2017xss,Aaij:2015bpa,ATLAS:2013cia,Khachatryan:2016csy}. 
At leading order in the QCD coupling constant, the pair creation of heavy quarks essentially produces them back-to-back in \pt. Experimentally, substantial deviations from this scenario are observed, which renders angular correlations between HF particles a sensitive probe of higher-order production mechanisms, such as gluon radiation or gluon splitting into a $Q\overline{Q}$ pair~\cite{Andronic:2015wma}. Future precision data on angular distributions are important to constrain the various contributions, thereby providing critical input to accurately evaluate and interpret hot-medium effects for HF measurements in A+A collisions.
\end{textbox}




\section{Heavy-quark diffusion in QCD matter}
\label{sec:diff}
The large masses of heavy quarks delay their thermalization in a QCD medium relative to light constituents, parameterically by a factor of $m_Q/T$. This delay renders the HQ thermalization time comparable to the fireball lifetime in URHICs, and therefore the modifications of HQ spectra carry sensitive information on their coupling strength to the expanding medium. Furthermore, since the HQ mass is large compared to typical temperatures, the propagation of low-momentum heavy quarks through the medium bears close analogy to ''Brownian motion", characterized by (many) elastic collisions with the medium of comparatively small momentum transfer. The pertinent transport framework is that of the Fokker-Planck equation for the HQ distribution function, $f_Q$, schematically written as
\beq
\frac{\partial}{\partial t} f_Q(t,p) = \frac{\partial}{\partial p} p A(p) f_Q(t,p)+ \frac{\partial^2}{\partial^2\vec{p}} B(p) f_Q(t,p) \ .
\eeq
The fundamental medium properties are encoded in temperature- and momentum-dependent transport coefficients $A$ and $B$, representing the thermal relaxation rate and momentum-diffusion of the heavy quark, respectively. The pertinent spatial diffusion coefficient, 
\begin{equation}
{\cal D}_s = \frac{T}{m_Q A(p=0)} \ , 
\end{equation}
characterizes the long-wavelength limit of HF transport, and as such is similar in nature to the electric conductivity or shear viscosity (characterizing electric charge and energy-momentum transport, respectively).

The basic observables to analyze HQ diffusion in URHICs, as introduced in Eqs.~(\ref{eq:raa}) and (\ref{eq:v2}), are the nuclear modification factor, \RAA, and elliptic flow, $v_2$ of charm and bottom hadrons and their decay products (such as semi-leptonic decay electrons or muons). Due to the relatively small momentum transfers, $q^2\sim T^2 \ll p_Q^2 \sim m_Q T$, imparted on a heavy quark, many collisions with the medium constituents are necessary to pick up the collectivity of the expanding fireball (this is quite different from energy loss at high momentum where often only one interaction per event occurs). The tell-tale signatures are: (a) a characteristic ``flow bump" in the \RAA, at a transverse momentum reflecting the typical expansion velocity of the medium, $p_T \sim m_Q \gamma_\perp v_\perp$ (which in practice is modified by thermal motion, incomplete thermalization and the denominator in the \RAA), and (b) a large $v_2$ which provides the possibly most direct gauge of the interaction strength (until reaching the values of the thermal limit). Note that for flow velocities of, say, $v_\perp=0.6c$, a $D$-meson at rest in the local rest frame carries a lab momentum of $\sim$1.5\,GeV/$c$, illustrating that low-momentum $D$-meson observables directly probe the charm diffusion coefficient. 

In the remainder of this section, we compile and discuss recent $D$-meson $R_{\rm AA}$ and $v_2$ data (Sec.~\ref{ssec_raa-v2}), followed by a brief discussion of the theoretical implications for the extraction of the HF diffusion coefficient (Sec.~\ref{ssec_diff-theo}).

\subsection{Collective flow of $D$-mesons} 
\label{ssec_raa-v2}

\begin{figure}[th]
\centering
\begin{minipage}[t]{0.49\linewidth}
\includegraphics[width=1\textwidth]{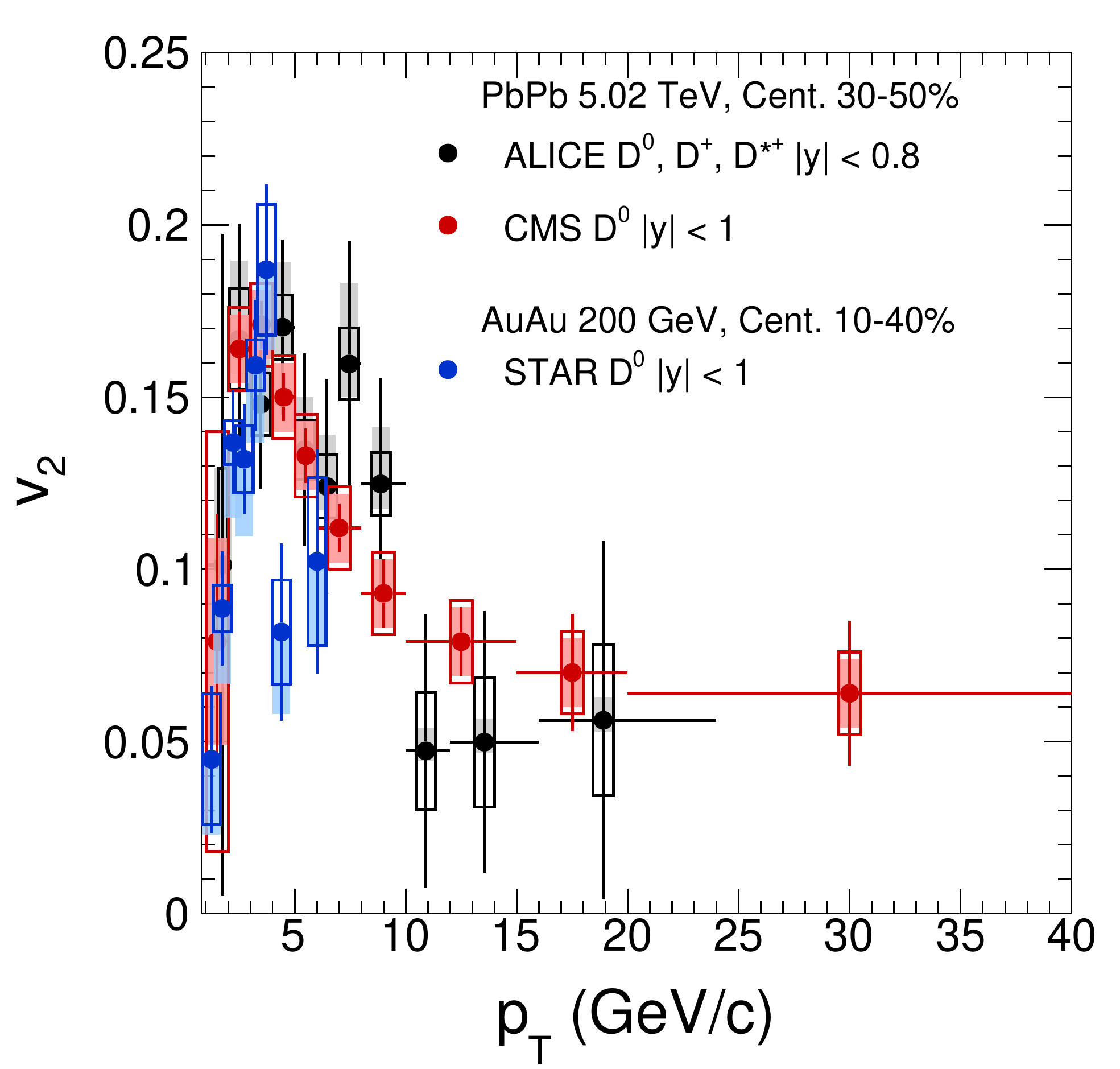}
\end{minipage}
\begin{minipage}[t]{0.49\linewidth}
\includegraphics[width=1\textwidth]{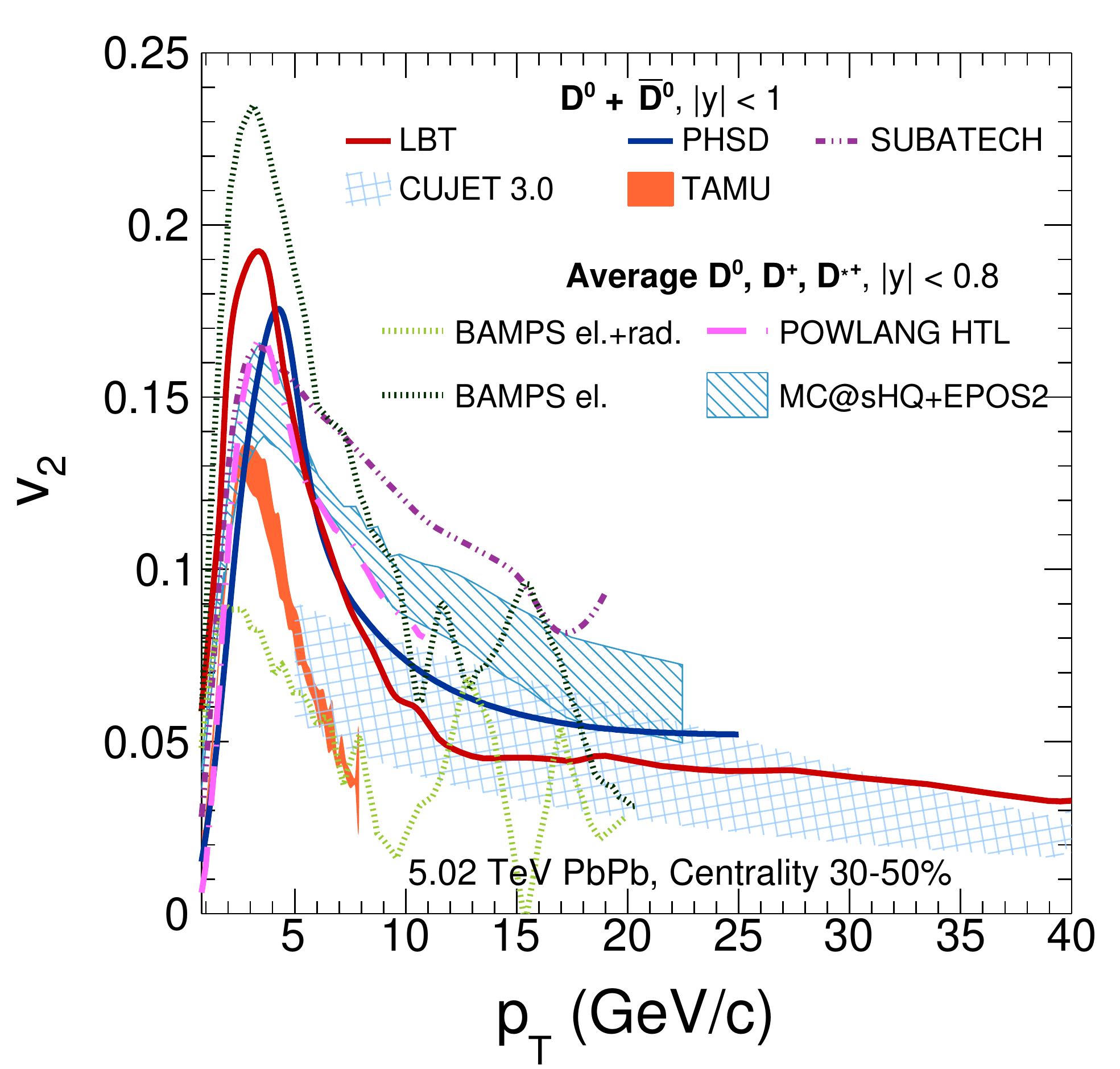}
\end{minipage}
\caption{Elliptic flow of $D$-mesons: (Left) experimental data in 30--50\% Pb+Pb (\sqrtsnn\,=\,5.02\,TeV) collisions by ALICE~\cite{Acharya:2018hre} and CMS~\cite{Sirunyan:2017plt} at the LHC, and in 10--40\% Au+Au (\sqrtsnn\,=\,200\,GeV) collisions by STAR~\cite{Adamczyk:2017xur} at RHIC; (Right) theoretical calculations for 30--50\% Pb+Pb (\sqrtsnn\,=\,5.02\,TeV) collisions ~\cite{Song:2015sfa,Djordjevic:2015hra,Xu:2015bbz,Xu:2014ica,Kang:2014xsa,Cao:2017hhk,Song:2015ykw,Horowitz:2015dta,Uphoff:2014hza,Nahrgang:2013xaa,Beraudo:2014boa, Nahrgang:2014vza}.} 
\label{fig:v2-AA}
\end{figure}

\begin{figure}[th]
\centering
\begin{minipage}[t]{0.48\linewidth}
\includegraphics[width=1\textwidth]{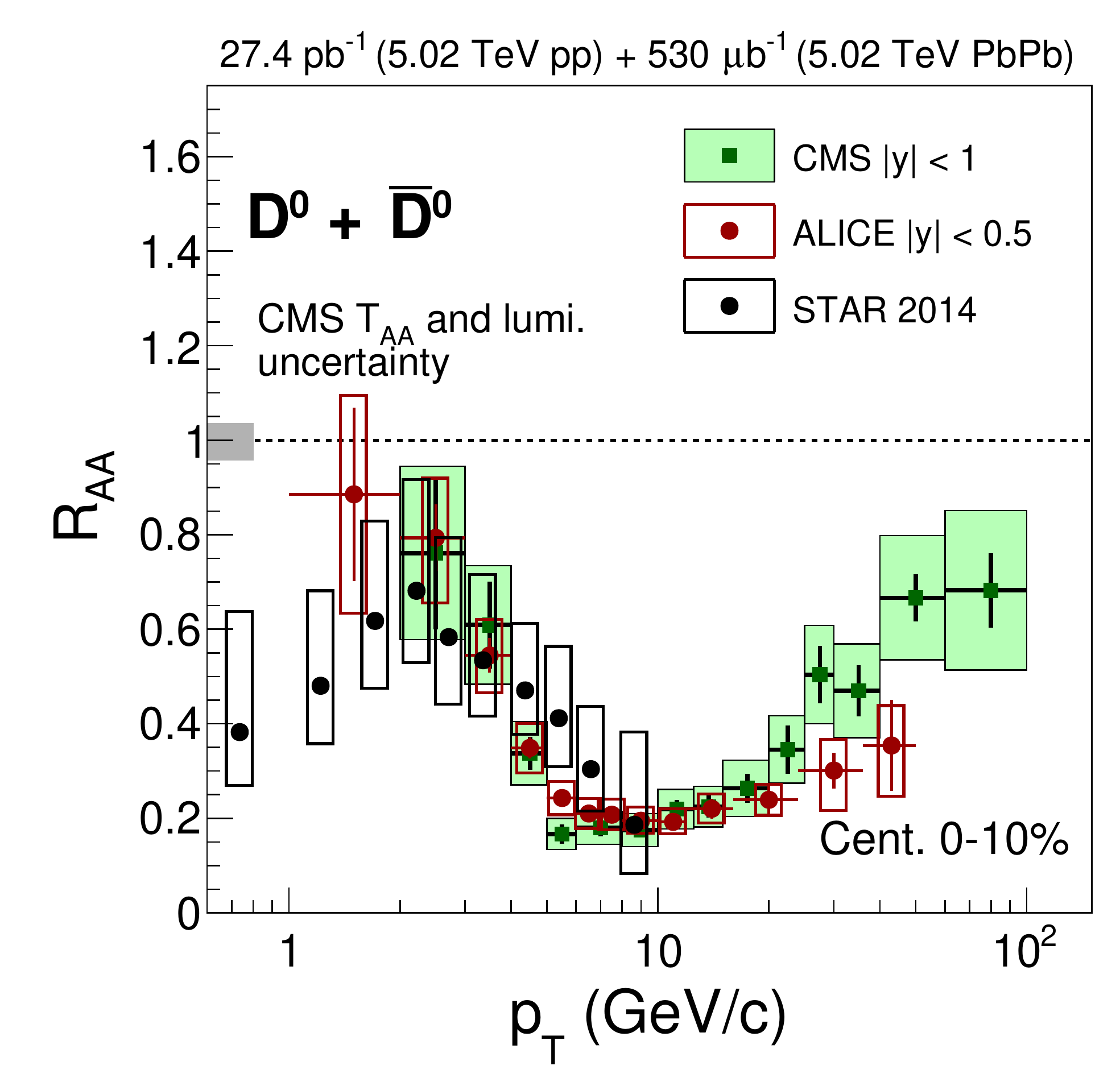}
\end{minipage}
\begin{minipage}[t]{0.48\linewidth}
\includegraphics[width=1\textwidth]{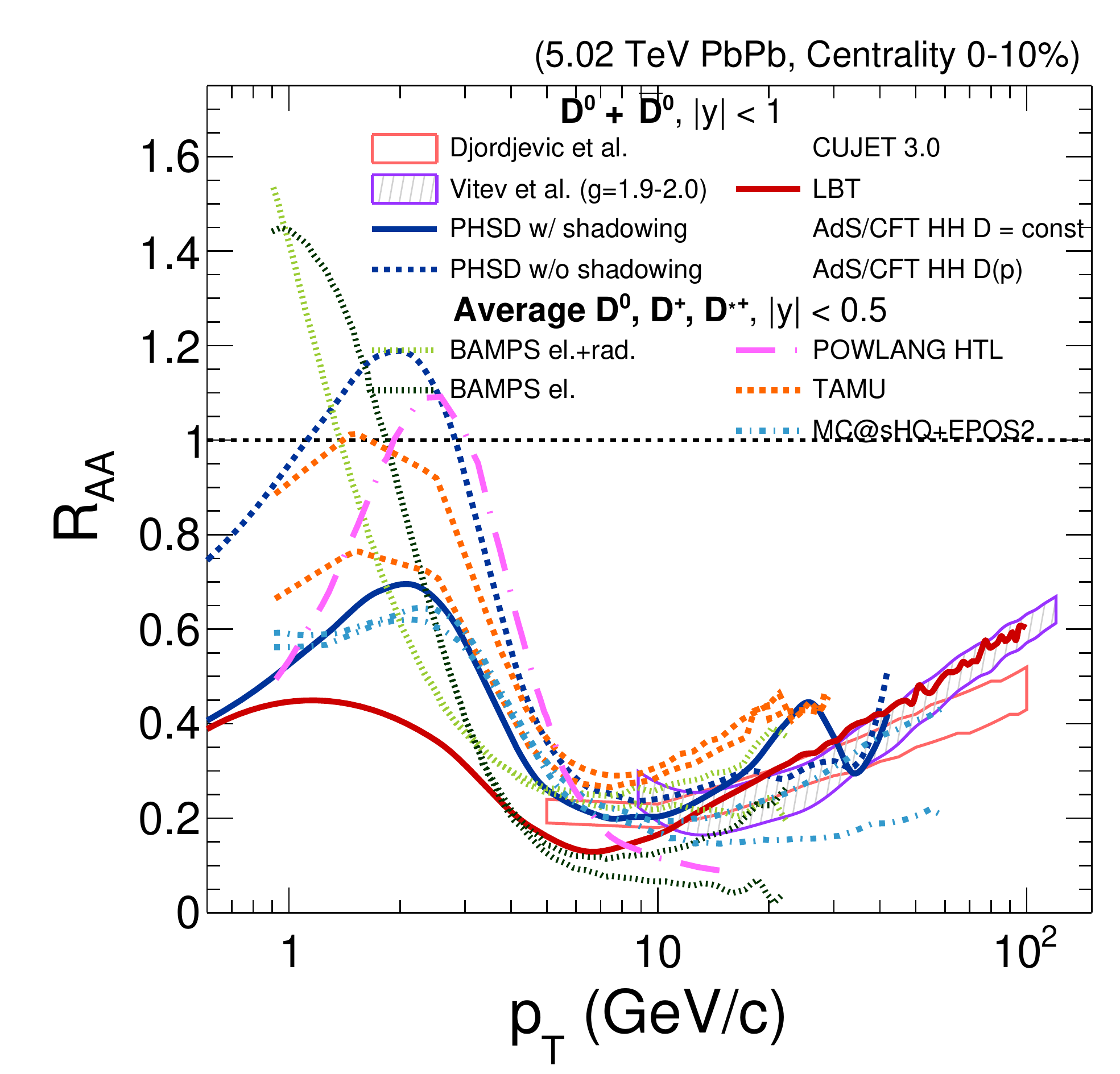}
\end{minipage}
\caption{(Left) Nuclear modification factor of $D$-mesons: (Left) experimental data in 0--10\% Pb+Pb (\sqrtsnn\,=\,5.02\,TeV) collisions by ALICE~\cite{Acharya:2018hre} and CMS~\cite{Sirunyan:2017xss} at the LHC, and in 0--10\% Au+Au (\sqrtsnn\,=\,200\,GeV) collisions by STAR~\cite{Adam:2018inb} at RHIC; (Right) theoretical calculations for 0--10\% Pb+Pb (\sqrtsnn\,=\,5.02\,TeV) collisions~\cite{Song:2015sfa,Djordjevic:2015hra,Xu:2015bbz,Xu:2014ica,Kang:2014xsa,Cao:2017hhk,Song:2015ykw,Horowitz:2015dta,Uphoff:2014hza,Nahrgang:2013xaa,Beraudo:2014boa,Chien:2015vja,Cao:2016gvr}.}
\label{fig:RAA}
\end{figure}

In Figs.~\ref{fig:v2-AA} and~\ref{fig:RAA} the left panels summarize recent $D$-meson $v_2$ and $R_{\rm AA}$ data in (semi-)central A+A collisions at RHIC ($\sqrt{s_{_{\rm NN}}}$\,=\,200\,GeV)~\cite{Adamczyk:2017xur,Adam:2018inb} and the LHC ($\sqrt{s_{_{\rm NN}}}$\,=\,5.02\,TeV)~\cite{Sirunyan:2017xss,Acharya:2018hre,Sirunyan:2017plt}; for comparison, the right panels give an overview of various model calculations of these quantities in Pb+Pb collisions at the LHC. 

At both collision energies large values of the $v_2$ are observed, peaking above 15\% at $p_T\simeq$ 3\,GeV/$c$, decreasing thereafter and leveling off around $\sim$5\% for $\pt\gsim10$\,GeV/$c$. A similarly close agreement, within current uncertainties, is also found for the \RAA data, both between ALICE and CMS as well as between LHC and RHIC data. This is not trivial given the differences caused by the factor of $\sim$25 difference in collision energy: \eg, the primordial \pt spectra of heavy quarks in $p$+$p$ collisions are much steeper at RHIC than at the LHC, and the initial medium temperature is expected to be significantly higher at the LHC. Future data to be taken with the upgraded ALICE detector and high-statistics CMS capabilities beyond LHC run-2~\cite{Citron:2018lsq}, and with the sPHENIX detector beyond 2022~\cite{Adare:2015kwa}, will greatly improve the data precision especially below $\pt\simeq5$\,GeV/$c$, which will allow to shed more light on the apparent agreements at the different energies.

The pronounced peak in the $v_2$ at low \pt, together with the corresponding maximum structure observed in the low-$p_T$ \RAA, provides clear evidence for an appreciable collectivity of $D$-mesons in URHICs. This conclusion is further corroborated in that the magnitude of the $D$-meson $v_2$ goes hand-in-hand with the one observed in the bulk medium (\ie, in the light-flavor sector) when varying, \eg, the collision centrality or system size. Most of the theoretical calculations shown in the right panels of Figs.~\ref{fig:v2-AA} and ~\ref{fig:RAA} are consistent with these features, although they differ in detail. For example, the height of the flow bump in the \RAA directly depends on the amount of nuclear shadowing that is included in the charm production cross section, as is specifically illustrated in calculations by the TAMU~\cite{He:2014cla} (dashed orange curves, including a shadowing range of 64-76\% and a coalescence uncertainty) and PHSD groups~\cite{Song:2015ykw} (dashed vs. solid dark-blue curves representing no or EPS09 shadowing~\cite{Eskola:2009uj}, respectively); it also depends
on the hadro-chemistry of charm-hadron production at the confinement transition, \eg, through enhanced $D_s^+/D$ and $\Lambda_c^+/D$ ratios relative to $p$+$p$ collisions. To utilize the location of the flow bump as measure of the radial flow picked up by the charm quarks and hadrons, an improved control over the bulk evolution models and the hadronization mechanisms will be necessary~\cite{Rapp:2018qla}. The magnitude of the $v_2$ is therefore a more direct measure of the low-momentum coupling strength, approaching the equilibrium limit from below. Its implications on the HF diffusion coefficient will be discussed in the next section. Another noteworthy feature of the $v_2$ data, also reflected in most theoretical calculations, is its rather pronounced peak at relatively low $p_T$, followed by a drop-off at intermediate $p_T\simeq$\,5-10\,GeV/$c$ and a leveling off around 5\% for $p_T\gsim$\,10\,GeV/$c$. Similarly, one can identify three corresponding regions in the \RAA: a maximum structure at low $p_T$, bottoming out near 10\,GeV/$c$ and then slowly rising thereafter. These structures suggest a strong-coupling regime with many rescatterings leading to large collectivity for $p_T\lsim$\,5\,GeV/$c$, a kinetic transition regime with few scatterings for 5\,GeV/$c$\,$\lsim p_T\lsim$\,10\,GeV/$c$, and an energy loss regime with radiative interactions for $p_T\gsim$\,10\,GeV/$c$ where path length differences generate a relatively small azimuthal asymmetry. 



\subsection{Implications for the heavy-quark diffusion coefficient}
\label{ssec_diff-theo}
\begin{figure}[th]
\includegraphics[width=3.5in]{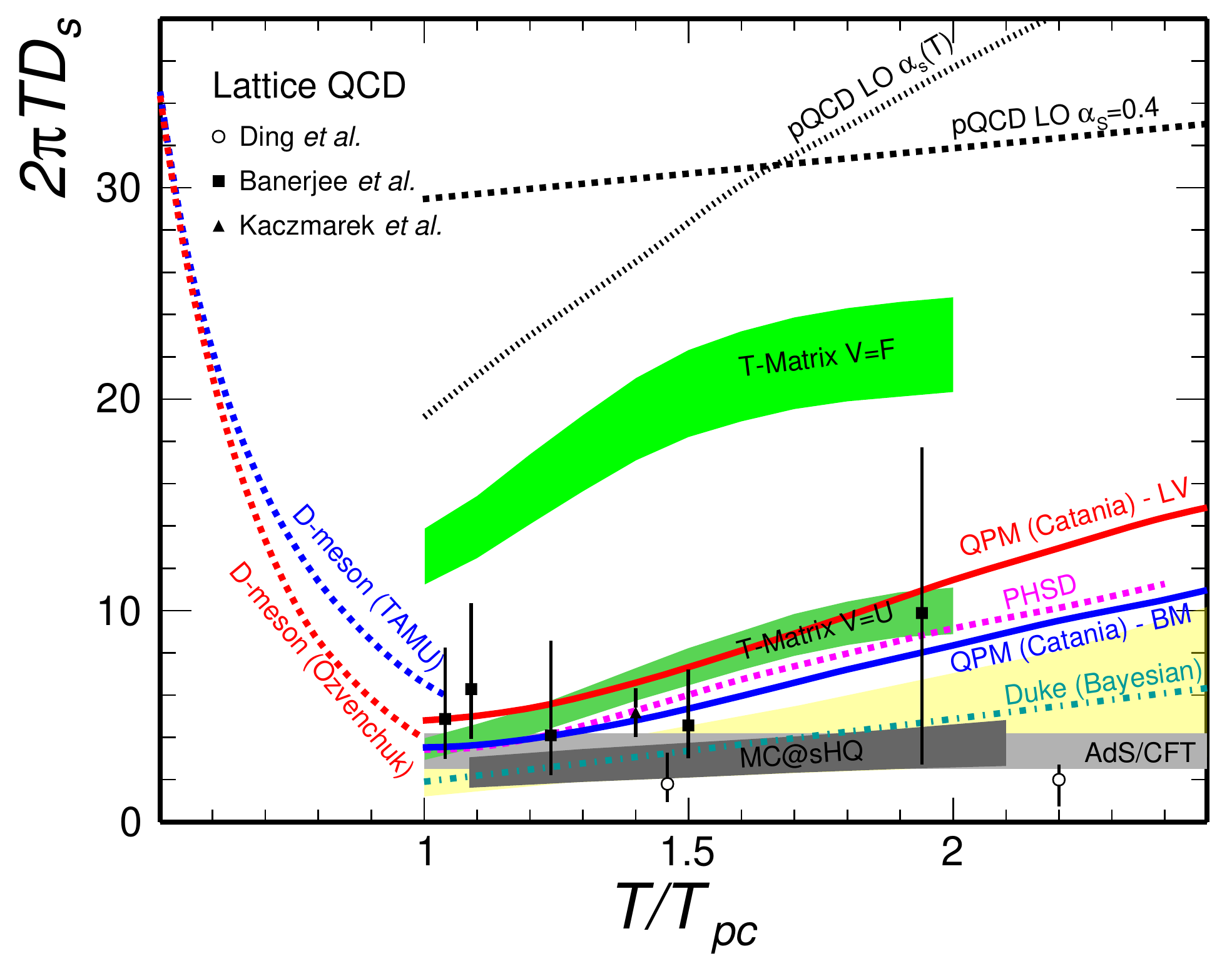}
\caption{Charm-quark spatial diffusion coefficient, $2\pi T{\cal D}_s$, as a function of reduced temperature $T/T_{\rm pc}$~\cite{Dong:2019unq}. Recent results from quenched lQCD (squares~\cite{Banerjee:2011ra}, open circles~\cite{Ding:2012sp} and triangles~\cite{Kaczmarek:2014jga}) are compared to model calculations based on different elastic interactions in the QGP: LO pQCD~\cite{Moore:2004tg,vanHees:2004gq} (dashed and dotted black lines), QPM calculations~\cite{Das:2015ana} utilizing Boltzmann- (BM; solid blue line) or Langevin-based (LV; solid red line) extractions, a dynamical QPM as utilized in PHSD~\cite{Song:2015sfa} (magenta dotted line), $T$-matrix approach with free energy ($F$) or internal energy ($U$) as potential~\cite{Riek:2010fk} (light- and dark-green bands), MC@sHQ perturbative approach with running coupling~\cite{Andronic:2015wma} (dark-grey band), and AdS/CFT-based calculations~\cite{Horowitz:2015dta} (light-grey band). The yellow band shows a Bayesian analysis fit result (90\% C.R.) using the Duke hydro/transport model~\cite{Xu:2017obm}. Also shown are the results for $D$-meson diffusion below \Tpc~using effective hadronic interactions~\cite{He:2011yi,Tolos:2013kva} (blue- and red-dashed lines, respectively).
}
\label{fig:Ds}
\end{figure}
A recent compilation of calculations of the dimensionless scaled HF diffusion coefficient, $2\pi T{\cal D}_s$, as a function of the reduced temperature $T/T_{\rm pc}$ is shown in Fig.~\ref{fig:Ds}~\cite{Dong:2019unq}. The overlap with the calculations shown for the \RAA and $v_2$ in the right panels consists of the PHSD model (a microscopic transport model using elastic HQ interactions in both QGP and hadronic phase), the Nantes model (MC@sHQ using an ideal-hydro evolution (EPOS-2) with both elastic and radiative interactions in a pQCD framework with running coupling and reduced Debye mass), and the TAMU model (non-perturbative elastic $T$-matrix interactions based on the internal-energy potential in an ideal-hydro evolution). The following observations may be made. The TAMU model, with $2\pi T {\cal D}_s$ increasing from 3--4 at \Tpc~to 8--10 at 2\,\Tpc~\cite{Riek:2010fk}, does not generate enough $v_2$, implying an upper limit of the transport coefficient. The ${\cal D}_s$ used in the PHSD approach~\cite{Song:2015sfa} is slightly lower, but an up to 30\% larger $v_2$ is generated. Part of this increase is presumably due to larger effects in the hadronic phase, and possibly due to differences in the bulk evolution (which are more difficult to discern). The Nantes approach requires a significantly smaller ${\cal D}_s$ than PHSD and TAMU. This is in large part due to the inclusion of radiative interactions, which are less effective in generating elliptic flow, since the collectivity of the incoming medium particle gets distributed over a 3-particle final state. This feature is also seen in the pQCD-based BAMPS parton cascade~\cite{Uphoff:2014hza}, where an elastic-only scenario (with retuned coupling constant) generates a much larger $D$-meson elliptic flow than the combined elastic+radiative one. An interesting recent finding reported in Ref.~\cite{Gossiaux:2019hp} is that when switching from an ideal-hydro evolution (in EPOS-2) to a viscous one (in EPOS-3), larger ${\cal D}_s$ values are extracted (of up to 20--30\%), presumably due to the delay in cooling caused by a local reheating of the expanding medium. This would, \eg, allow to improve the TAMU calculations of the $v_2$ which currently underestimate the data. Quasiparticle models (QPMs), as used, \eg, by the Catania group~\cite{Das:2015ana} , also suggest $2\pi T {\cal D}_s$ values with a minimum of around 4 near \Tpc, gradually rising thereafter. On the other hand, a Bayesian analysis performed by the Duke group~\cite{Xu:2017obm} with a functional ansatz for temperature and momentum dependences of the transport parameters, extracts lower values near 2, also rising with $T$, while the AdS/CFT approach is, by construction, $T$-independent. Current lQCD data are restricted to computations in quenched approximation, \ie, without dynamical quark loops; they are generally consistent with model calculations that yield a good phenomenology. 

In summary, current phenomenology suggests a HQ diffusion coefficient in a range of $2\pi T {\cal D}_s$ = 2--4 near $T_{\rm pc}$, consistent with previous estimates~\cite{Rapp:2018qla,Prino:2016cni}. Leading-order perturbative results~\cite{Svetitsky:1987gq,CaronHuot:2007gq} are ruled out, and even non-perturbative calculations with a potential close to the HQ free energy extracted from lQCD, are not viable. Most of the calculations show an increasing trend with temperature, but such a trend has not been firmly established from data yet. However, it has been argued~\cite{Das:2015ana,Rapp:2008zq} that the correlation between the $v_2$ and \RAA indeed prefers an increasing coupling strength as the temperature decreases toward \Tpc, essentially to generate a large $v_2$ (which is most effective in the later stages when most of the bulk medium $v_2$ has built up) while not over-suppressing the \RAA (which mostly occurs in the early densest phases of the fireball); recent Bayesian data analysis~\cite{Xu:2017obm} also find such a trend. These arguments, however, require a good control over the \pt dependence of the transport coefficients. General arguments of a decreasing coupling strength with increasing temperature (due to a reduced screening length) and momentum (due to asymptotic freedom) are in line with the behavior of most model calculations. 


\section{Heavy-quark hadronization}
\label{sec:hadron}

The hadronization of quarks and gluons into color-neutral hadrons as observed in experiments is a fundamental process in QCD which, due to its nonperturbative nature, remains a challenging problem to date. Various models and parameterizations of hadronization processes have been developed in phenomenological studies of particle production in elementary collisions. For high-$p_T$ partons, fragmentation functions in connection with collinearly-factorized production processes have been successfully applied to describe a wide range of measured hadron spectra. These functions depend on the flavor of the primordial parton but are usually assumed to be universal and as such can be constrained using the $e^+$+$e^-$ or $e^-$+$p$ collision data and tested in hadronic collisions. As the $p_T$ is lowered, non-universal effects appear. Flavor asymmetries in forward-rapidity production of strange- and charm-meson production in $\pi$+$p$, $p$+$p$, $\pi$+A and $p$+A collisions have been observed and attributed to coalescence processes with comoving valence quarks, see, \eg, Ref.~\cite{Das:1977cp}. Color reconnection schemes~\cite{Sjostrand:1993hi} have been implemented into event generators to supplement fragmentation processes. Alternative to the microscopic description for low-$p_T$ hadron production, statistical models have been put forward; they produce fair fits to light- and strange-flavor hadron production in elementary collisions with a common hadronization ``temperature" of $T_H\simeq$\,160\,MeV~\cite{Becattini:1997uf}, with some caveats (such as extra suppression factors or correlation volumes)~\cite{Andronic:2008ev,Kraus:2008fh}. 

In heavy-ion collisions novel features in the production systematics of hadrons have emerged. As mentioned above, the low-\pt spectra ($\pt\lsim$\,2--3\,GeV/$c$) are well described by a locally thermalized fireball whose hadro-chemistry is frozen near the phase transition temperature, \Tpc, whereas kinetic freezeout (where elastic rescattering ceases) occurs at a lower temperature of about $T_{\rm fo}\simeq100$\,MeV, with a large collective-flow velocity of up to $\sim$\,0.9\,$c$ near the surface. In the intermediate-\pt region ($\sim$3--6\,GeV/$c$) a marked 
baryon-to-meson enhancement, relative to $p$+$p$ collisions, has been observed. In principle, this could be a remnant of collective-flow effects (although with incomplete thermalization), pushing heavier particles further out in \pt than light ones. Alternatively, quark coalescence models, where hadrons are formed via recombination of nearby constituent quarks at the phase boundary (with a smaller collective flow than at kinetic freezeout), have been proposed and successfully applied to reproduce the $p/\pi$ and $\Lambda/K_{\rm S}$ ratios in this \pt  region~\cite{Lin:2003jy,Fries:2008hs}. The coalescence models can simultaneously reproduce the so-called constituent-quark number scaling (CQNS) of the elliptic flow, which deviates from the hydrodynamic behavior at intermediate \pt. Interestingly, CQNS works better at RHIC than at the LHC, possibly due to stronger rescattering effects in the hadronic phase at higher collision energies.

Since heavy quarks are predominantly produced in initial hard scatterings, they can serve as a tag in the hadronization process and illuminate underlying mechanisms as outlined above. In addition, their incomplete thermalization extends the sensitivity to different hadronization mechanisms down to low \pt, in a spirit similar to the intermediate-\pt regime for light hadrons. Thus, HF hadro-chemistry is a unique tool to probe the fundamental issue of hadronization. For example, coalescence models have predicted the ratio of strangeness carrying $D_s$-mesons over non-strange $D$-mesons to be enhanced in A+A relative to $p$+$p$ collisions, 
as a consequence of the enhanced strangeness production in the former~\cite{Kuznetsova:2006bh,He:2012df}. In analogy to the light sector~\cite{Sorensen:2005sm,MartinezGarcia:2007hf}, they furthermore predict an enhancement in the charm-baryon-to-meson ratio, \eg, $\Lambda_c/D^0$~\cite{Oh:2009zj}. The statistical hadronization model~\cite{Andronic:2003zv}, based on thermal production ratios at fixed charm number, makes qualitatively similar predictions, albeit quantitatively different. As will be discussed further below in Sec.~\ref{sec:qgp}, these two approaches are not necessarily exclusive of each other. Similar considerations apply in the bottom sector.

In the following, we discuss the current experimental status and possible interpretations in the production of strangeness-carrying charm and bottom mesons (Sec.~\ref{ssec:Ds}), as well as HF baryons (Sec.~\ref{ssec:Lc}).

\subsection{Strange heavy-flavor mesons}
\label{ssec:Ds}
The ground state charm-strange meson, $D_s^+(1968)$, has been extensively measured in elementary collisions. Its production ratio with respect to $D^0$ mesons, $D_s^+/D^0$, has been well constrained to be around 0.15--0.20~\cite{Lisovyi:2016qjn}, with a slight increase as a function of $p_T$. Various event generators (\eg, PYTHIA) have been tuned as to reproduce this ratio in $p$+$p$ collisions.

\begin{figure}[!t]
\centering
   \begin{subfigure}[b]{0.45\textwidth}
    \includegraphics[width=\textwidth]{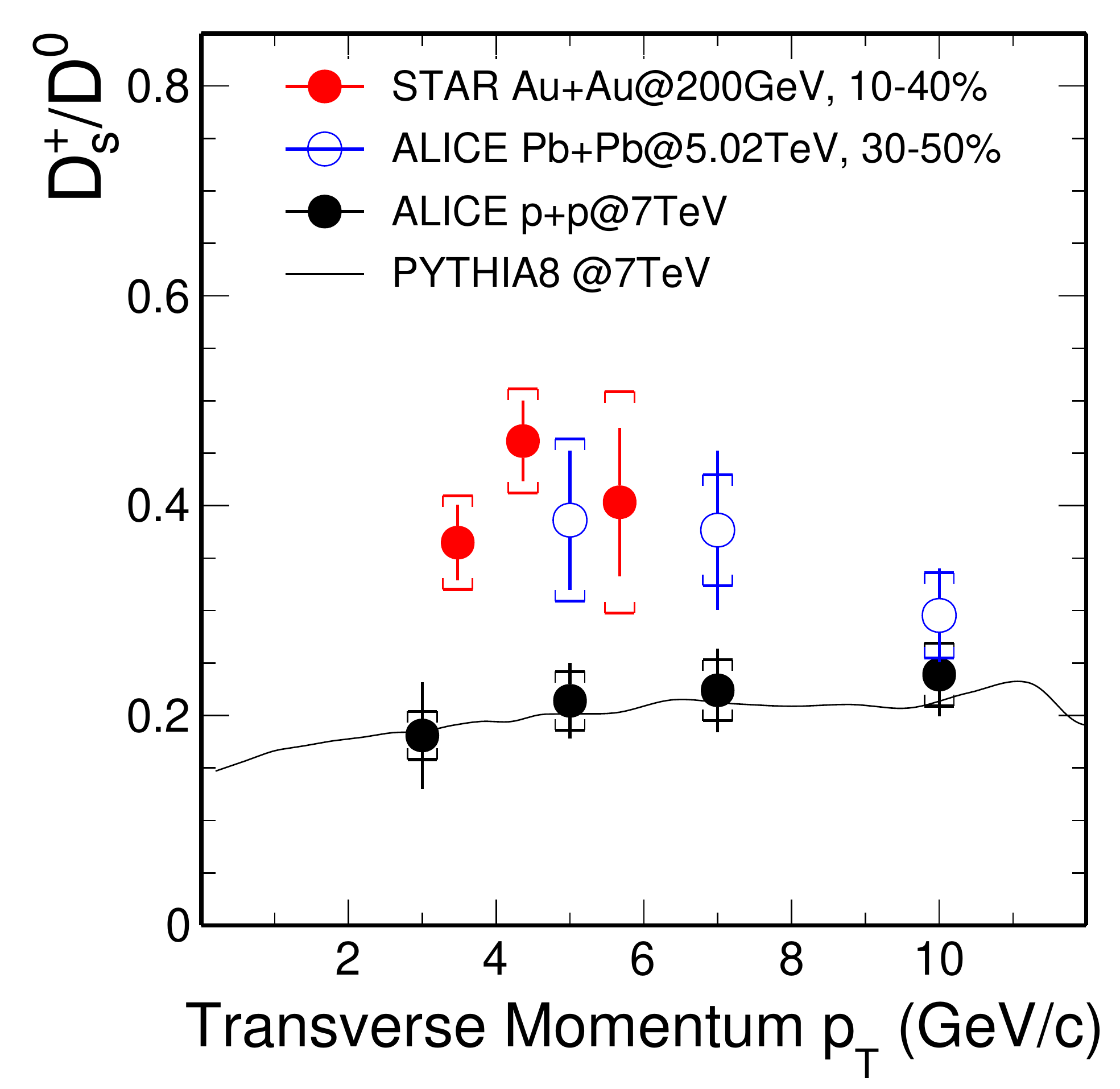}
  \end{subfigure}
  \begin{subfigure}[b]{0.48\textwidth}
    \includegraphics[width=\textwidth]{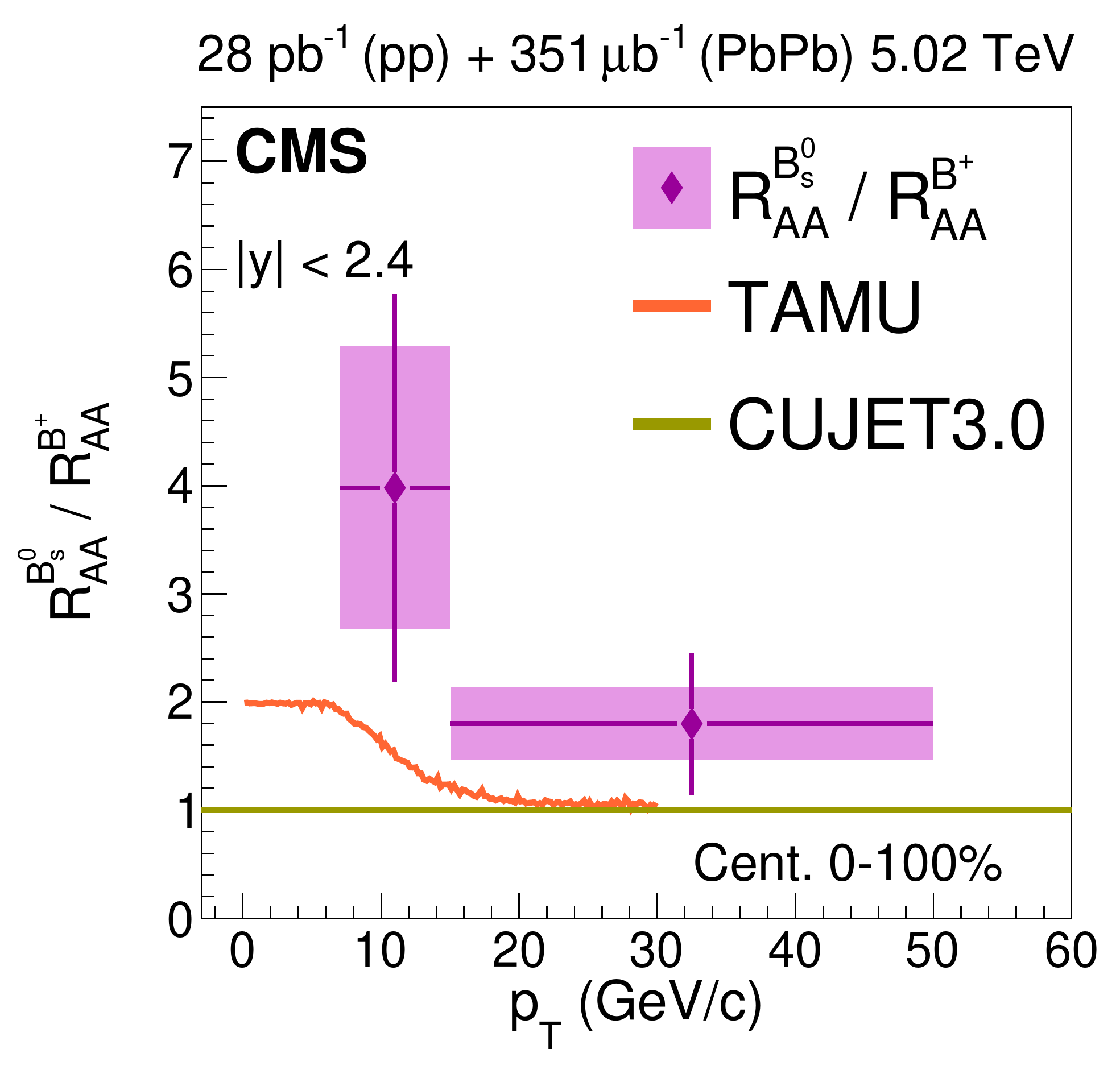}
  \end{subfigure} \\
  \begin{subfigure}[b]{0.8\textwidth}
    \includegraphics[width=\textwidth]{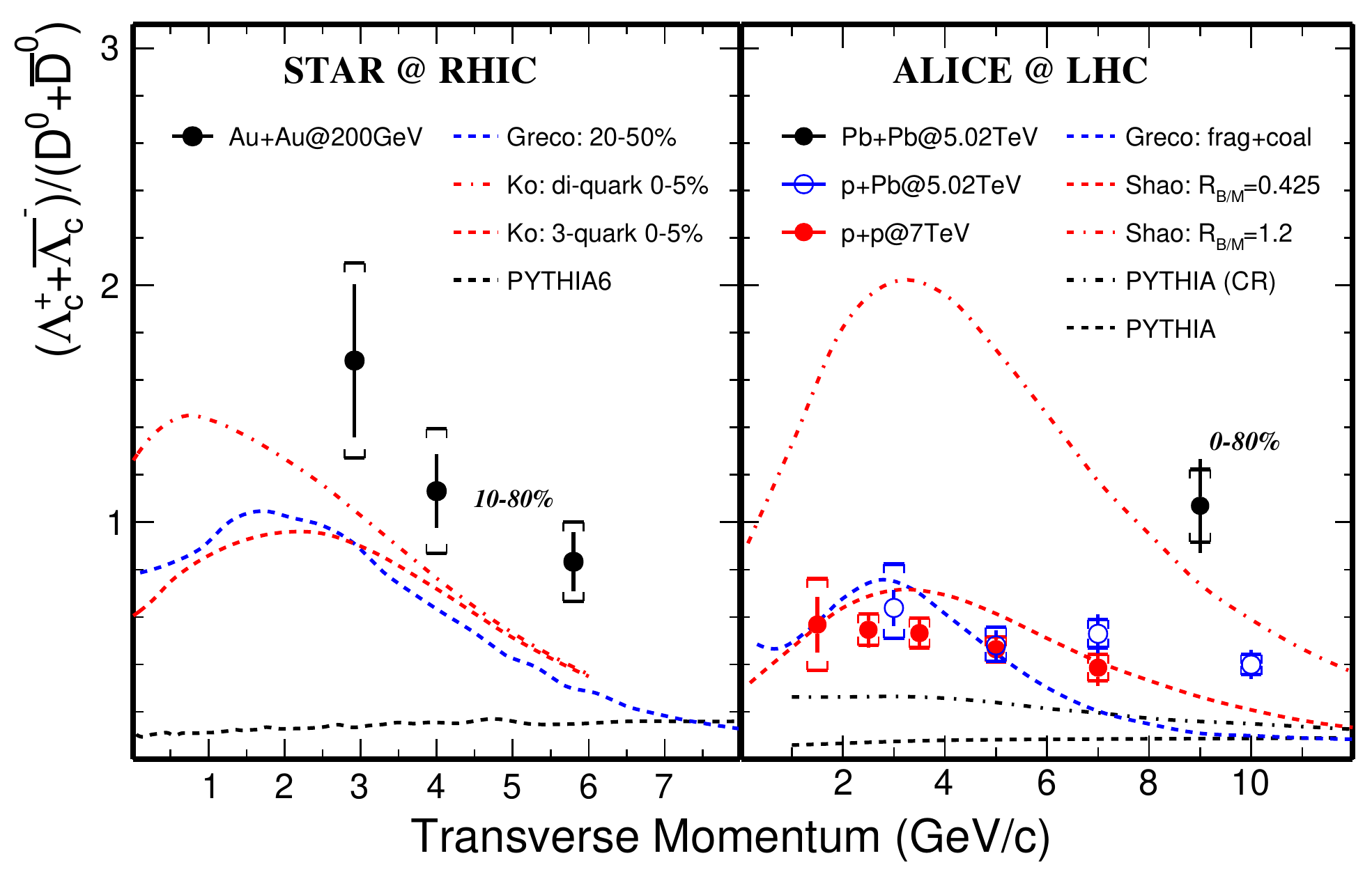}
  \end{subfigure}
\caption{(Upper Left) $D_s^+/D^0$ ratios measured in 10--40\% Au+Au (\sqrtsnn\,=\,200\,GeV) collisions by STAR~\cite{STARDsLc}, 30--50\% Pb+Pb (\sqrtsnn\,=\,5.02\,TeV) and $p$+$p$ (\sqrts\,=\,7\,TeV) collisions by ALICE~\cite{Acharya:2018hre}. (Upper Right) $R_{\rm AA}$ double-ratio of $B_s^0$ and $B^+$ in 0--100\% Pb+Pb (\sqrtsnn\,=\,5.02\,TeV) collisions from CMS~\cite{Sirunyan:2018zys}. (Lower) $\Lambda_c^+/D^0$ ratios as a function of $p_{T}$ measured by STAR in Au+Au (\sqrtsnn\,=\,200\,GeV)~\cite{STARLcQM18} (left sub-panel) and by ALICE in $p$+$p$, $p$+Pb and Pb+Pb collisions~\cite{Acharya:2018ckj} (right sub-panel). Data points are compared to various model calculations (see text) for heavy-ion collisions, as well as PYTHIA calculations.}
\label{fig:DsLcD0}
\end{figure}

Recent measurements of $D_s^+/D^0$ ratios in A+A collisions at RHIC~\cite{STARDsLc} and the LHC~\cite{Acharya:2018hre} are highlighted in the upper left panel of Fig.~\ref{fig:DsLcD0} and compared to that in $p$+$p$ collisions at the LHC~\cite{Acharya:2017jgo}. The latter are well reproduced by a PYTHIA-8 calculation (solid curve). In semi-/central heavy-ion collisions, a significant enhancement of this ratio is found, which is compatible between RHIC and LHC energies in the overlapping \pt region. 
The observed values are consistent with the prediction of $\sim$0.35 from the statistical hadronization model. The transport results by the TAMU group~\cite{He:2012df}, including recombination at the hadronization transition, are also consistent with the data up to $p_T\simeq$\,4--5\,GeV/$c$, but fall below at higher \pt where the data are still enhanced above the $p$+$p$ value. Thus, recombination appears to be at work, but improved model calculations are needed to better describe the transition region from the near-equilibrium values to the fragmentation regime. 
Recent results from the CMS collaboration show a hint of an enhanced $B_s^0/B^+$ ratio in Pb+Pb collisions~\cite{Sirunyan:2018zys} with respect to $p$+$p$ collisions, as shown in the upper right panel of Fig.~\ref{fig:DsLcD0}. This is also qualitatively consistent with the expectation from coalescence, together with the strangeness enhancement in the QGP. More precise measurements and further theory developments will offer new insights on hadronization in the bottom sector.
 
\subsection{Heavy-flavor baryons}
\label{ssec:Lc}
The $\Lambda_c^+(2286)$ is the lowest-mass charm baryon. Its largest hadronic decay branching of $\sim$6.3\% is into the $pK^-\pi^+$ final state. Together with a short lifetime of $c\tau\sim 60\mu m$, the hadronic-decay reconstruction with three daughters makes it very challenging to measure in heavy-ion collisions. Recent high-luminosity runs together with precision silicon detectors allowed to perform first measurements in different collision systems. The lower panel of Fig.~\ref{fig:DsLcD0} shows the $\Lambda_c^+/D^0$ ratio as a function of $p_{\rm T}$ measured at RHIC~\cite{STARLcQM18} and the LHC~\cite{Acharya:2018ckj} in $p$+$p$, $p$+A and A+A collisions, compared to different model calculations. 

The first surprise comes in the elementary collision systems: the ALICE and LHCb measurements of the $\Lambda_c^+/D^0$ ratio in $p$+$p$ and $p$+Pb collisions~\cite{Acharya:2017kfy,Aaij:2018iyy} are much larger than the expected fragmentation baseline in default PYTHIA calculations. Recent developments using hadronization via color reconnection in the PYTHIA model or color ropes in the DEPSY model can increase the calculated $\Lambda_c^+/D^0$ ratio in the low- to intermediate-\pt regions. However, these calculations involve large uncertainties due to the choices of parameters implemented in the modified hadronization schemes. The relation of the physics underlying the color reconnection and rope hadronization schemes in $p$+$p$ collisions requires further investigations. In addition, even in $p$+$p$ collisions, hadronization through coalescence processes with surrounding valence (or sea) quarks may occur, as was briefly alluded to at the beginning of this section. Finally, the SHM predicts a $\Lambda_c^+/D^0$ ratio of about 0.2, well below the measured values of around $\sim0.5\pm0.1$. In all models feeddown from decays of excited states to the ground state plays an important role. While this is a well-defined contribution in the SHM (only requiring the masses and degeneracies of the excited states, plus measured branching ratios), it is less straightforward how resonances are populated in more microscopic descriptions. Progress on these questions will much benefit from more precise but also broader measurements including excited HF baryons in elementary collision systems.

The $\Lambda_c^+/D^0$ ratios measured in Au+Au and Pb+Pb collisions exhibit yet another significant enhancement of a factor of 2 or more over the $p$+$p$ measurements. The STAR data suggest that the enhancement further increases toward lower \pt, while the ratio may trend toward the $p$+$p$ values at high \pt. Model calculations using charm-quark coalescence can reach a qualitative agreement with the data~\cite{Plumari:2017ntm,Oh:2009zj,Li:2017zuj,Zhao:2018jlw}. The calculations invoke non-trivial wavefunction effects, such as diquark sub-structures, to produce a good part of the enhancement. Since this, in principle, implies different wavefunction parameters for different hadrons, the pertinent theoretical uncertainties are rather large. Also here, broader and more precise data will go a long way toward constraining the models, probing their approximations and/or offer new insights.

The large $\Lambda_c/D^0$ and $D_s/D^0$ ratios observed in experiment indicate that charm baryons and charm-strange mesons give sizable contributions to the total charm cross section. Complementary to that, the STAR measurement of the \pt-integrated \RAA for $D^0$ mesons gives only about 0.5 in central Au+Au collisions, \ie, the $D^0$ yield amounts to only 50\% of the expectation from the $N_{\rm bin}$ scaled cross section in $p$+$p$ collisions. One reason for such a suppression could be the reduction of the total charm cross section due to nuclear shadowing of the parton distribution functions in the incoming nuclei. However, at RHIC energies, this is not expected to be a large effect. On the other hand, the STAR collaboration found that, based on their measurements of all major charm-hadron ground states with suitable extrapolations to low $p_T$, the enhancement in $D_s$ and especially $\Lambda_c$ production in Au+Au collisions leads to a total charm cross section which is compatible with $N_{\rm bin}$ scaling from $p$+$p$ collisions. However, the study is currently conducted in a wide centrality bin, and the dominant uncertainty in heavy-ion collisions arises from a limited \pt coverage in the $\Lambda_c$ measurement. Future more precise measurements of charm baryons and charm-strange mesons are expected to much better quantify the total charm cross section at mid-rapidity, which also has important consequences, \eg, for the regeneration yield of charmonia~\cite{Rapp:2017chc}.  



\section{Heavy-quark energy loss in QGP}
\label{sec:eloss}
When hard-scattered partons, produced in collisions of heavy nuclei at high energy, pass through the subsequently formed QGP, they lose energy inside the medium~\cite{Bjorken:1982tu}. This parton energy loss, often referred to as ``jet-quenching", was first discovered at RHIC via a suppression of the \RAA (as defined in Eq.~(\ref{eq:raa}) of light hadrons at high \pt~\cite{Adams:2005dq,Adcox:2004mh,Back:2004je,Arsene:2004fa}, and confirmed at the LHC~\cite{Chatrchyan:2011sx,Aad:2010bu,Aamodt:2010jd}. These measurements enabled a first estimate of the jet transport coefficient (or “scattering power”), $\qhat$, of the QGP~\cite{Burke:2013yra}. 
Theoretical studies~\cite{Dokshitzer:2001zm,Braaten:1991we,Djordjevic:2003zk,Armesto:2003jh} have suggested the flavor dependence of jet-quenching as a fruitful testing ground for mechanisms underlying parton energy loss models. On the one hand, gluons have a larger color charge than quarks and are therefore expected to lose more energy when passing through the QGP. On the other hand, compared to light quarks, the scattering kinematics of heavy quarks implies that they are less likely to radiate off energy when passing through the medium, which in particular, manifests itself as a suppression of gluons emitted at angles smaller than the ratio of the quark mass to its energy, $\Theta \lsim m_Q/E$~\cite{Dokshitzer:2001zm}. This ``dead cone'' effect (and its disappearance at high \pt) can be studied by comparing the suppression patterns of HF hadrons relative to light-flavor ones in heavy-ion collisions. 

In the language of pQCD, there are two energy loss mechanisms: (1) collisional energy loss through elastic scatterings of the high-momentum heavy quarks off generally low-momentum medium particles; (2) energy loss via the radiation of gluons triggered by the acceleration due to the color forces between heavy quarks and the medium particles. The radiative energy loss in QCD is similar to the electromagnetic radiation off an accelerated charge, but with the essential difference that the radiated gluon itself can rescatter off the medium. The comparison of the magnitude of the heavy- and light-flavor energy loss could in particular illuminate the relative partition of the collisional and radiative contributions to the energy loss of partons when varying the \pt of the parent partons.

\subsection{Heavy-quark energy loss mechanism}
Since heavy quarks are not directly accessible experimentally, nuclear modification factors of HF mesons, non-prompt $J/\psi$ (from $b$-quark decays) and HF decay leptons have been measured at RHIC and the LHC over a wide \pt range in order to reveal the effects of their energy loss in the medium. In this section we focus on the high-\pt part in the compilation of the current status of experimental data and theoretical calculations for $D$ mesons shown in Fig.~\ref{fig:RAA}. 

The effect of energy loss of high-momentum heavy quarks plowing through the QGP will cause the HQ momentum spectra to ``shift" toward smaller values and thus lower the \RAA. At intermediate $p_T\simeq$\,5--10\,GeV/$c$, one presumably goes through a transition region, from low-\pt ``Brownian motion" (where the near-thermalized $c$-quarks accumulate, cf.~Sec.~\ref{ssec_raa-v2}) to effects of energy loss prevalent at high \pt. In the intermediate-\pt interval, the experimental data are of high accuracy and agree well between ALICE and CMS. Although the central values of the STAR data are above the LHC data, they are still compatible within their uncertainties. The \RAA data exhibit a rather pronounced drop and reach a minimum around 5--8\,GeV/$c$, which is reproduced by most model calculations shown in the right panel of Fig.~\ref{fig:RAA}. In analogy to the energy loss of charged particles in an electromagnetic medium through photon radiation, one expects gluon radiation to become the dominant mechanism for high-momentum partons in the QGP. Beyond the basic features in the HF-meson \pt spectra one has to rely on data-to-model comparisons to extract more quantitative information. 
For instance, in the BAMPS calculations (a pQCD-based parton transport model with 3-body interactions) the results for the \RAA with and without radiative energy loss (where the latter has been retuned with a $K$ factor to reproduce the $v_2$ in the light-hadron sector) show a rather pronounced difference in the intermediate-\pt region, with the former (latter) at the upper (lower) end of the data.  
In the TAMU calculation, which includes only elastic energy loss (without $K$ factor), a significant deviation from the data sets in for $\pt\gsim$\,5\,GeV/$c$, suggesting the onset of radiative contributions in this regime. 
On the other hand, the PHSD transport calculations, which also use only elastic interactions, give a good description of the high-\pt \RAA, with appreciable contributions from hadronic rescattering. 

At high $\pt> 10$\,GeV/$c$, radiative energy loss is expected to be the dominant interaction. The experimental \RAA values start rising, and continue to do so up to the highest \pt measured, reaching $\sim 0.7$ at 100\,GeV/$c$ in the CMS data. 
Perturbative-QCD model calculations (Soft Collinear Effective Theory (SCET$_{\rm G}$)~\cite{Chien:2015vja}, Djordjevic et al.~\cite{Djordjevic:2015hra} and Linear Boltzmann Transport (LBT)~\cite{Cao:2017hhk}) are consistent with the magnitude and the rise of the high-\pt \RAA data, mostly as a result of an interplay with the flattening HQ \pt spectral slope at production (\ie, in the denominator of the \RAA), and due to a reduced coupling at high \pt. However, some of the ingredients in these models, \eg, the bulk medium evolution or the value of the strong coupling constant, differ considerably~\cite{Rapp:2018qla,Cao:2018ews}. The agreement with pQCD calculations reiterates the challenge of understanding the transition into the low-\pt regime which is dominated by non-perturbative interactions, also calls for a detailed assessment of their differing ingredients.    

\subsection{Mass hierarchy of energy loss}
As pointed out in Sec.~\ref{sec:intro}, measurements of HQ energy loss serve as important tests on jet quenching models due to the expected mass hierarchy in gluon radiation, which, at a given \pt, is expected to be suppressed with increasing parton mass. The standard procedure is to compare the \RAA's of charm and bottom mesons (or their decay products, such as leptons or non-prompt $J/\psi$'s), and to light hadrons. 

\begin{figure}[!thb]
\centering
\begin{minipage}[t]{0.465\linewidth}
\includegraphics[width=1\textwidth]{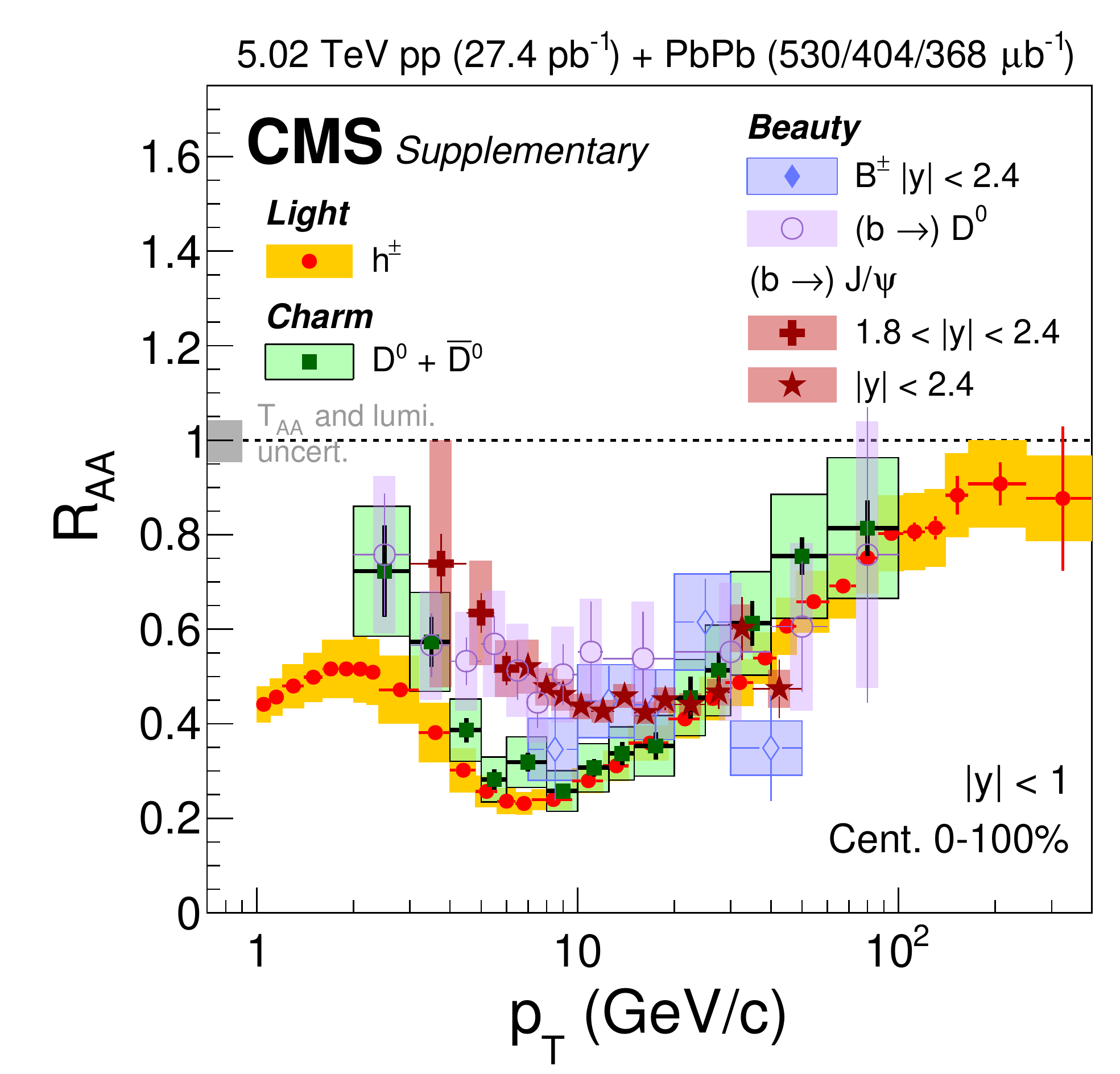}
\end{minipage}
\begin{minipage}[t]{0.525\linewidth}
\includegraphics[width=1\textwidth]{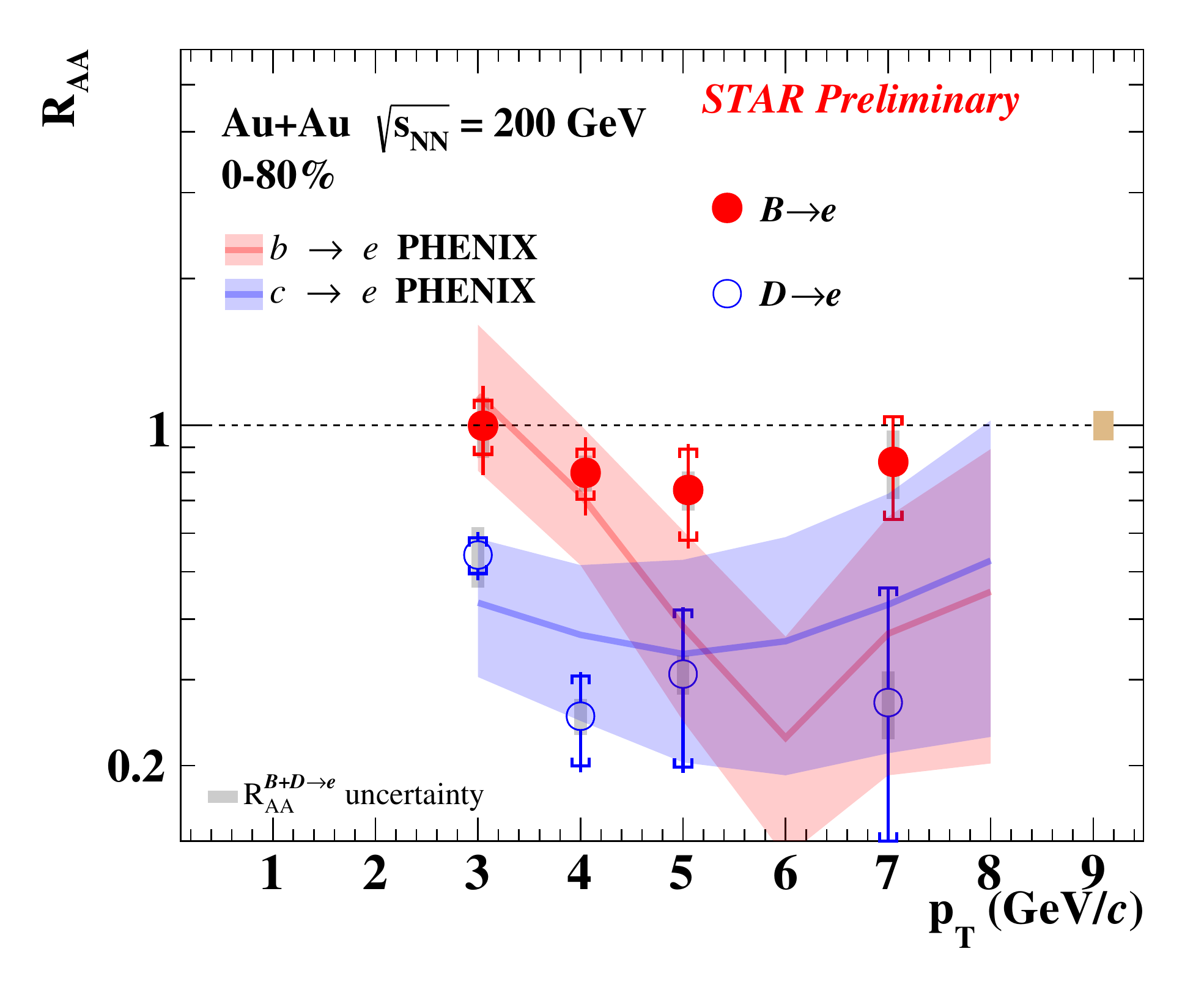}
\end{minipage}
\caption{(Left) Nuclear modification factors \RAA as a function of \pt in 0--100\% Pb+Pb collisions at $\sqrt{s_{_{\rm NN}}}$\,=\,5.02\,TeV collisions for neutral $D$ mesons (green squares) compared to inclusive charged hadrons (red dots), $B^\pm$ mesons (blue dots), non-prompt $J/\psi$ mesons (red crosses and stars) and non-prompt $D^0$ (open circles) from CMS~\cite{Sirunyan:2018ktu}. (Right) STAR (dots) and PHENIX (solid lines with bands) data for semileptonic HF decay electrons separated into bottom (red) and charm (blue) decays in minimum bias Au+Au collisions at $\sqrt{s_{_{\rm NN}}}$\,=\,200\,GeV~\cite{STARB,Adare:2015hla}.}
\label{fig:RAAvsFlavor}
\end{figure}

Such a comparison is compiled in Fig.~\ref{fig:RAAvsFlavor} for the \RAA's measured at the LHC (left panel) for charged particles ($h^\pm$)~\cite{Khachatryan:2016odn}, prompt $D^0$~\cite{Sirunyan:2017xss}, non-prompt $D^0$ (from $b\rightarrow D^0$ decays)~\cite{Sirunyan:2018ktu}, non-prompt $J/\psi$ (from $b\rightarrow J/\psi$ decays)~\cite{Sirunyan:2017isk}, and $B^\pm$~\cite{Sirunyan:2017oug}, and at RHIC (right panel) for semi-leptonic decay electrons which have been separated in bottom ($b\rightarrow e$) and charm contributions ($c\rightarrow e$)~\cite{STARB,Adare:2015hla}. At intermediate \pt, the non-prompt $J/\psi$ \RAA is significantly higher than the $D^0$ \RAA indicating a bottom-charm hierarchy. Since the non-prompt $J/\psi$ does not carry the full \pt of the parent $B$ meson, a horizontal shift of its \RAA to slightly higher \pt (by around 2\,GeV/$c$ based on {\sc pythia+evtgen}-based studies) should be accounted for in a more direct comparison to the prompt $D^0$'s (which reinforces the difference). The recently measured non-prompt $D^0$ \RAA is consistent with the observations made with non-prompt $J/\psi$'s, corroborating the evidence for a flavor hierarchy. Indications for a similar hierarchy are observed at RHIC from HF decay electrons reported by STAR and PHENIX, although experimental uncertainties are still rather large. At very high \pt, when the energy-to-mass ratio (Lorentz-$\gamma$ factor) becomes large, one expects the mass effects to cease and the bottom, charm and light-flavor \RAA's to degenerate (with the caveat of gluon fragmentation contributions due to their different color charge). The CMS data are consistent with such a scenario, with a charm-light degeneracy in the \RAA's for $\pt\gsim$\,10\,GeV/$c$, and a further degeneracy with bottom for $\pt\gsim$\,25\,GeV/$c$ or so. If confirmed with higher-precision data, this would support the dominance of radiative energy loss for charm and bottom quarks beyond the minimum in their \RAA, as well as the factor $\sim$3 difference in the onset of the degeneracy expected from the bottom-to-charm mass ratio, \ie, at essentially the same $\gamma$ factor.

While the quark mass hierarchy observed in the pertinent \RAA measurement at intermediate \pt is very suggestive, quantitative conclusions about the nature of the energy loss as a function of parton flavor require careful comparisons between data and model calculations. For example, the primordial spectral \pt slopes of charm and bottom can be quite different, and the partition of elastic and radiative interactions in the transition regime leading up to the minimum of the \RAA's can be quite different, too (and, in fact, also at high \pt). For instance, in the Djordjevic model, the heavy- and light-flavor data are consistently described if the mass effects are included in the calculations, even though the coherence effects in gluon emission, which generate a nontrivial path length dependence, are found to degenerate only near $\pt\simeq100$\,GeV/$c$~\cite{Djordjevic:2018ita}. The size of other effects, related to HF meson decays (non-prompt $J/\psi$, $D^0$ and semileptonic decays), nuclear parton distribution functions or HQ hadronization are still elusive at this point. For the latter two, in the case of heavy quarks, one might hope that they approximately scale with their mass ratio (for the same Bjorken-$x$ in their production, or the same velocity ($\gamma$ factor) relative to the medium which determines their coalescence probability).

Experiments at the LHC have also attempted to measure $b$- and $c$-tagged jet \RAA's~\cite{Chatrchyan:2013exa,ALICED0jet}. Limited by the experimental accuracy, the interpretation of the physics underlying these measurements is not yet clear. Future high-statistics data analyses with upgraded detectors beyond Run-2 at the LHC~\cite{Citron:2018lsq} and with sPHENIX at RHIC~\cite{Adare:2015kwa} are expected to greatly improve the data accuracy.

\subsection{Heavy-flavor triggered correlations}
As an additional discrimination power of elastic vs.~radiative interactions of heavy quarks, beyond the \RAA and $v_2$ , angular correlation measurements between two heavy quarks have been proposed~\cite{Gossiaux:2006yu,Zhu:2007ne,Akamatsu:2009ya,Cao:2015cba}. This type of correlation is different from HF elliptic flow measurements, which are correlations between heavy- and light-flavor mesons. The idea is to test different deflection patterns, such as multiple low-momentum transfer elastic scatterings vs.~few relatively large-momentum transfer interactions (radiative energy loss).

Furthermore, high-momentum probes (of various flavors) can serve as projectiles in QGP scattering experiments which aim at resolving its short-distance structure. By analyzing the outgoing particles from those experiments, including both medium particles and possibly the attenuated projectiles, one hopes to gain insights about the QGP at varying wavelength~\cite{DEramo:2018eoy}. Once again, the advantage of HF particles is the ability to track their flavor content as they probe the medium. By measuring the modifications of the angular correlation functions between light- and heavy-flavor particles, one hopes to study how the nearly perfect QCD fluid emerges from quarks and gluons.

These types of measurements have recently been commenced as $D$-$\overline{D}$, $D^0$-hadron and $D^0$-jet angular correlation analyses. Currently, the former two are limited to $p$+$p$ collisions~\cite{Ma:2017sel,ALICE:2016clc}. The results can be described reasonably well by event generators such as PYTHIA6 and PYTHIA8, which can then serve as a baseline for the future measurements with high-statistics A+A data. CMS has reported the first $D^0$-jet angular correlation measurement in $p$+$p$ and Pb+Pb collisions at $\sqrtsnn= 5.02$\,TeV~\cite{CMS:2018ovh}.
By comparing the angular displacement profiles of the $D^0$ relative to the jet axis in $p$+$p$ and Pb+Pb collisions, an indication of a larger displacement between the $D^0$ and the dominant energy flow in the jet has been reported. This is consistent with the expectations from HQ diffusion; future measurements with significantly better accuracy can be performed with high luminosity LHC data~\cite{Citron:2018lsq}.

\section{How can heavy flavor probe the "inner workings" of QCD matter?}
\label{sec:qgp}
The understanding of the spectral and transport properties of heavy flavor is an important objective of heavy-ion collisions in its own right, but a core premise was the idea that HF particles can {\em probe} the medium, \ie, unravel spectral and transport properties of the bulk matter and contribute to understanding the microscopic mechanisms underlying them. Given the inherent scale dependence of QCD, one expects marked variations as the momentum of the probe is varied (note, however, that it is the momentum {\em transfer} from the probe to the medium that determines the resolution, much like in the Rutherford experiment; this means that even high-momentum probes can, in principle, still be sensitive to rather soft scatterings). Heavy quarks are an ideal tool for such an investigation due to their sensitivity to transport properties over the entire range of their momenta (light partons lose their sensitivity once they thermalize). In this section we discuss some of the ramifications that the insights from HF transport are yielding into the QGP medium. The discussion is organized into the long-wavelength properties probing the long-range forces and transport coefficients (Sec.~\ref{ssec_diff}), intermediate wavelengths probing hadronization and the transition from elastic to radiative interactions (Sec.~\ref{ssec_hadro}), and short wavelengths probing energy loss (Sec.~\ref{ssec_eloss}). 

\subsection{Long Wavelengths: Diffusion}
\label{ssec_diff}
The dimensionless scaled HQ diffusion coefficient, $2\pi T {\cal D}_s$, characterizes the (inverse) coupling strength of the QGP, \eg, ${\cal D}_s \sim 1/(\alpha_s^2T)$ in pQCD, up to logarithmic corrections~\cite{Svetitsky:1987gq}. As such it is expected to carry universal information; also note that the HQ mass dependence is ``divided out'' in its relation to the thermalization rate, eq.~(\ref{fig:Ds}), which is explicit in the perturbative calculation but has also been approximately found in nonperturbative calculations~\cite{Huggins:2012dj} (implying a universality of the spatial diffusion coefficient for charm and bottom ). 
After all, the microscopic interactions governing the transport of heavy flavor must also be operative in the transport of energy-momentum (encoded in viscosities) or electric charge (encoded in the conductivity), suggesting a proportionality of the pertinent dimensionless quantities ${\cal D}_s (2\pi T) \sim \eta/s \sim \sigma_{\rm EM} / T$~\cite{Rapp:2009my}. The (double-) ratio of these quantities is expected to acquire different values depending on the nature of the medium; \eg, in a weakly coupled QGP, one finds ${\cal D}_s (2\pi T) / (\eta/4\pi s)\simeq 2.5$, while for the strong-coupling limit in the gauge-gravity duality this ratio turns out to be one~\cite{Policastro:2002se}.  

\begin{figure}[!thb]
\centering
\begin{minipage}[t]{0.49\linewidth}
\includegraphics[width=0.95\textwidth]{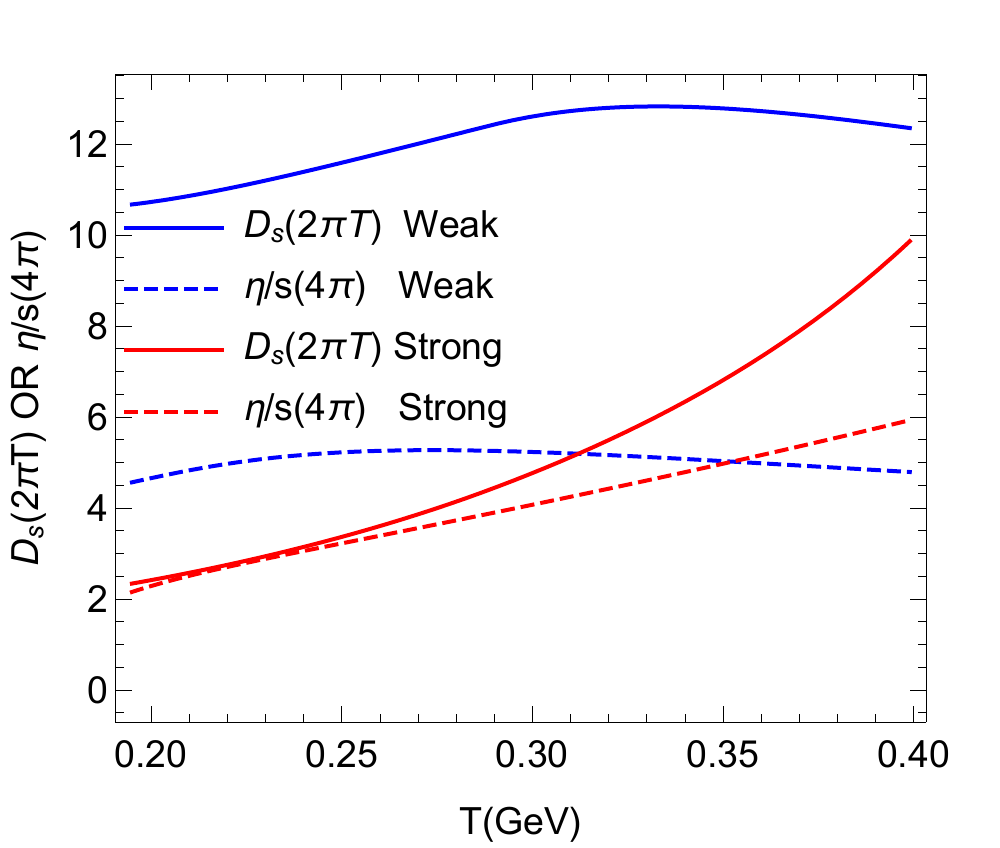}
\end{minipage}
\begin{minipage}[t]{0.49\linewidth}
\includegraphics[width=0.95\textwidth]{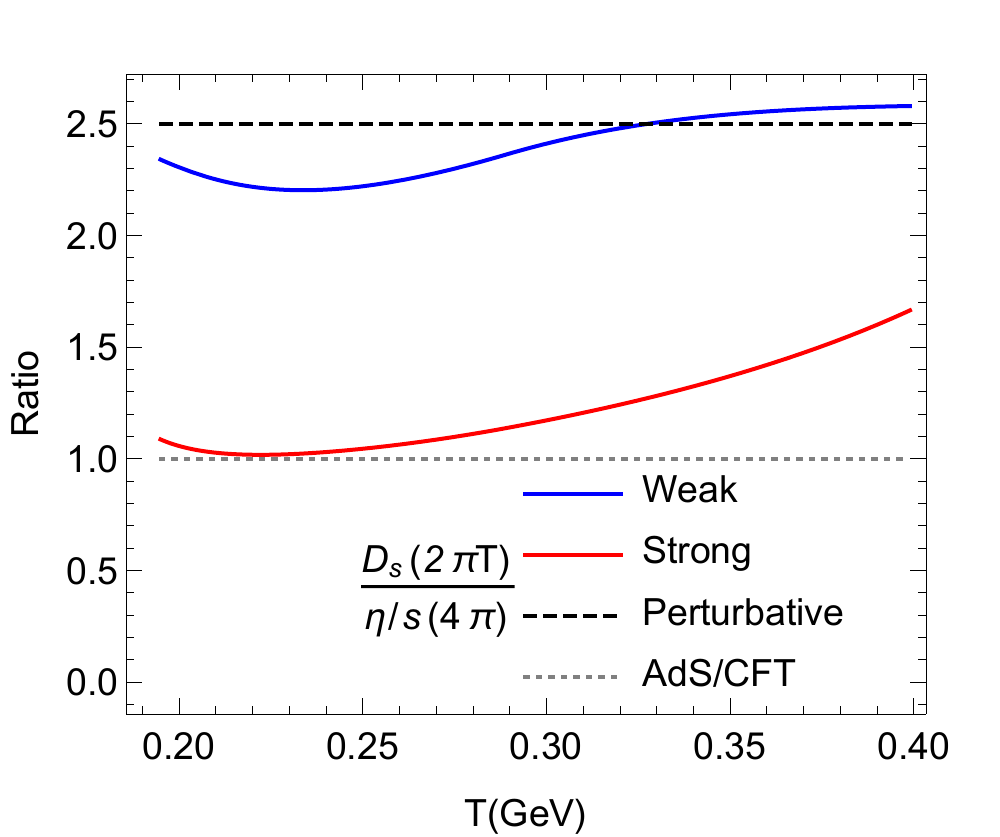}
\end{minipage}
\caption{Left: Comparison of the dimensionless-scaled HQ diffusion coefficient (\Ds$2\pi T$; solid lines) and specific shear viscosity ($\eta/s$; dashed lines) computed in a self-consistent many-body theory in two scenarios for the in-medium color force (blue lines: weakly coupled potential close to the HQ free energy; red lines: strongly coupled potential, well above the free energy)~\cite{Liu:2016ysz}. Right: ratio of the dimensionless HQ diffusion coefficient to specific shear viscosity for a strongly (red line) and weakly (blue line) coupled scenario, compared to a weakly coupled perturbative calculation (dashed line) and to the strong-coupling limit in gauge-gravity duality (dotted line).} 
\label{fig:Ds-eta}
\end{figure}

The current calculations and extractions of the HQ diffusion coefficient in the QGP require a large enhancement, by about an order of magnitude, over the baseline LO pQCD results (recall Fig.~\ref{fig:Ds}). While in practice this is implemented in different ways ($K$ factor, running coupling in Born diagrams, or HQ potential within $T$-matrix), the generic feature is a long-range component of the force between the diffusing quark and the QGP. A natural candidate is the remnant of the confining force above \Tpc. It is well established from lQCD computations that the would-be order parameter of confinement, the Polyakov loop, changes only rather gradually with temperature above \Tpc~\cite{Borsanyi:2010bp,Cheng:2006aj}. In the thermodynamic $T$-matrix approach, for example, remnants of the confining force are implemented via an in-medium Cornell potential constrained by lQCD data and turn out to be critical in generating small diffusion coefficient~\cite{Riek:2010fk}. Within the same approach, the ratio of shear viscosity over entropy density has been calculated~\cite{Liu:2016ysz}, see the left panel of Fig.~\ref{fig:Ds-eta}. Two different input potentials representing a weakly-coupled scenario (WCS; close to the HQ free energy) and a strongly coupled one (SCS; closer to the internal energy), which both reproduce the lQCD equation of state, give rather different results for the HQ diffusion coefficient: ${\cal D}_s$ differs by a factor of $\sim$\,5 close to \Tpc, while $\eta/s$ only differs by up to factor of 2. Interestingly, the ratio of the two transport coefficients in the two scenarios suggests that the WCS and the SCS are close to the pertinent weak- and strong-coupling limits, respectively. Since the HQ diffusion coefficient in the WCS is clearly too large for URHIC phenomenology, one concludes that QCD matter near \Tpc~is indeed a strongly coupled system. This is further corroborated by inspecting the underlying light-parton spectral functions: for the WCS, they exhibit rather well-defined quasi-particle peaks with widths of about $\Gamma_q$\,=\,0.1--0.2\,GeV, while in the SCS quark and gluon widths of 0.5--1\,GeV melt the low-momentum quasi-particle peaks near \Tpc. However, at higher momenta and temperatures~\cite{Liu:2017qah} well-defined quasi-particles re-emerge. The latter is also encoded in the rise of the double ratio of the transport coefficients in the right panel of Fig.~\ref{fig:Ds-eta}.



\subsection{Intermediate Wavelengths: Hadronization and Elastic vs. Radiative Interactions}
\label{ssec_hadro}
As discussed in Sec.~\ref{ssec_raa-v2}, the data for the $D$-meson \RAA and $v_2$ are suggestive for a transition from a collective to a kinetic regime of HF interactions in URHICs, for $p_T\simeq5\to10$\,GeV/$c$. If one can control the low-momentum (elastic) HQ interactions to a sufficient extent, one would obtain a quantitative handle to constrain the interplay of elastic and radiative interactions, in particular when also including $B$-meson observables. Even in the light-parton sector, elastic interactions may still play a significant role in the high-$p_T$ suppression of the single-inclusive hadron \RAA~\cite{Qin:2007rn}, which is not easily disentangled from radiation. 

Another context in which the intermediate-$p_T$ region is believed to contain valuable information is hadronization, specifically its modification in the environment of a QGP. As light quarks are no longer expected to be thermalized in this regime, their hadronization leaves more direct fingerprints on the produced hadron spectra and $v_2$. Coalescence processes are widely believed to be at the origin of the constituent-quark number scaling (CQNS) of the $v_2$ and the enhancement of $p/\pi$ or $\Lambda/K$ ratios observed at RHIC~\cite{Fries:2008hs}. However, CQNS is not accurately satisfied at the LHC. Extensions of coalescence model calculations to low $p_T$ need to satisfy energy conservation to ensure a proper matching to the (thermal and chemical) equilibrium limits. Such approaches have been developed~\cite{Ravagli:2007xx,Cassing:2008sv}. Systematic investigations of HF hadro-chemistry from low to intermediate $p_T$ therefore provide an excellent opportunity to improve the understanding of hadronization processes of the QGP. Again, the large mass of charm and bottom quarks is a critical ingredient as to preserve the ``identity" of the hadronizing quark (since pair production is strongly suppressed). From a microscopic perspective, hadronization processes involve confining interactions; at high \pt, these are usually modelled by fragmenting strings or ropes. At low \pt, however, the color-neutralizing interactions can occur with surrounding partons from the QGP, via a two-body scattering into (pre-) hadronic (or diquark) resonances. The natural candidate for this interaction is the linear (``string") term in the Cornell potential which gradually emerges as the temperature cools toward \Tpc. This is implemented, \eg, in the TAMU model where remnants of the confining potential are essential to generate a small HQ diffusion coefficient in the QGP near \Tpc. At the same time, this interaction leads to the formation of $D$-meson like resonances, which catalyze the hadronization process. The emerging confining force thus plays the dual role of generating a strongly coupling QGP and causing its hadronization (recall that lQCD computations of the so-called ``interaction" measure, $(\epsilon-3P)/T^4$, find it to peak just above \Tpc). In this sense, one can interpret the CQNS of the light-hadron $v_2$ at intermediate \pt as a manifestation of the strong coupling near \Tpc.

\subsection{Short Wavelengths: Energy Loss}
\label{ssec_eloss}
The commonly used transport coefficient associated with high-energy partons traversing
the QCD medium is the transverse broadening parameter, 
\begin{equation}
\hat{q} = \frac{\Delta p_T^2}{\lambda} = \frac{4D_p E_p}{p} \ .
\end{equation}
It is  directly related to the transverse-momentum diffusion coefficient, $D_p$, in the Fokker-Planck equation, and as such momentum dependent.  
Progressing from intermediate to high momenta, recombination effects in the hadronization process cease (as the probability to find comoving partons from the medium drops), and therefore a more pristine window on the energy loss mechanisms opens up. A mass dependence may survive to rather high $p_T$, in terms of a different partition of radiative and collisional processes between light and heavy quarks, and differences in the coherence of the radiation. Both effects affect the magnitude and path length dependence of the energy loss and thus the measured \RAA and $v_2$. For heavy flavor, a good control over the dominantly collisional energy loss in the low- and intermediate-$p_T$ regime seems to be within reach. If so, one obtains a handle on deciphering the partitioning of radiative and collisional contributions at high $p_T$, which would help to better understand the energy loss in the light-parton sector as well. Non-perturbative effects (such as remnants of the confining force) could still play a role in small-angle scattering (small $Q^2$) at high \pt. On the other hand, large-angle scattering, a short-wavelength probe, has been suggested to resolve the color-charged quasi-particles in the QGP~\cite{DEramo:2018eoy}. A new promising observable in this context is the angular diffusion of heavy flavor in a jet~\cite{CMS:2018ovh}, taking advantage of the unique combination of a reference axis (the axis of energy flow, with a well defined angular distribution as measured in $p$+$p$) and HF tagging.   

In a recent EMMI task force report~\cite{Rapp:2018qla}, a comparison of radiative energy loss calculations indicated a close-to-linear path length dependence for bottom quarks in all approaches for $p_T$ up to at least 40\,GeV/$c$. For charm quarks, more noticeable deviations set in for $p_T\gsim10$\,GeV/$c$, increasing with momentum. In addition, the magnitude of the energy loss in different perturbative implementations exhibits differences of several 10's of percent, even though the models are benchmarked against data, see also Ref.~\cite{Cao:2018ews}. A recent analysis of the path length ($L$) dependence based on comparisons of large (Pb+Pb) and medium-size (Xe+Xe) collision systems found rather moderate deviations from a linear behavior, and a relatively weak flavor dependence, over a large range in $p_T$~\cite{Djordjevic:2018ita}. Nevertheless, with increasing \pt from $\sim$20\,GeV/$c$ on, a stronger-than-linear dependence becomes more pronounced first for light flavors before merging with charm and bottom near 100\,GeV/$c$ at a power of $\sim$$L^{1.4}$. 



\section{Heavy-flavor production in small collision systems}
\label{sec:pA}

The original goal of measuring open HF production in $p/d$+A collisions is to study ``cold nuclear matter" (CNM) effects that include: (a) a modification of the nucleon parton distribution functions in nuclei which affects the primordial production of HF particles, \ie, their yields and spectra; 
(b) a nuclear broadening (``Cronin effect") or energy loss of the incoming partons when the nuclei penetrate each other, prior to the HF production process;
(c) final-state interactions of produced heavy quarks in the nuclear medium of the passing-by remnants of the incoming nucleus A. In practice, the distinction of these effects is not straightforward, as, \eg, their factorization is not easily established. Nevertheless, quantifying the size of the CNM effects is important for the interpretation of the A+A data where these effects are, in principle, also present and thus should be separated from the ``hot-medium" effects of the subsequently formed fireball. In the following we first discuss the phenomenology of CNM effects in $p/d$+A collisions (Sec.~\ref{ssec:cnm}) and then address the issue of possible hot-medium effects in these systems from the HF perspective (Sec.~\ref{ssec:pA-challenge}).


\subsection{Cold-nuclear-matter effects}
\label{ssec:cnm}
Measurements of charm and bottom mesons and their HF decay leptons at RHIC and the LHC show that their modifications in minimum-bias (MB) $p/d$+A collisions at mid-rapidity are relatively small~\cite{Adare:2012yxa,Abelev:2014hha,Khachatryan:2015uja,Adam:2016ich}. At low \pt the mid-rapidity $R_{p\rm A}$ at the LHC appears to be slightly suppressed, consistent with expectations from nuclear shadowing~\cite{Eskola:2009uj} (although the uncertainties are still rather large), while at higher \pt it is compatible with one. The charm-hadron and lepton $R_{p\rm A}$'s or $R_{d\rm A}$'s show a larger suppression at forward rapidity (\ie, in the $p$ or $d$ going direction probing small Bjorken-$x$ in the nucleus) at both RHIC and the LHC~\cite{Aaij:2018iyy,Adare:2013lkk,Aaij:2017gcy}, again compatible with nuclear shadowing, as well as gluon saturation models. The measurements of charm- and bottom-hadron and -jet $R_{p\rm A}$ in various rapidity windows at the LHC can be reasonably well described by models with either initial-state modifications or final-state CNM energy loss, due to the currently large experimental uncertainties~\cite{Khachatryan:2015uja,Khachatryan:2015sva,Sirunyan:2016fcs}. 
 
 A cleaner access to nuclear PDFs~\cite{Eskola:2016oht,Guzey:2013xba} may be obtained from recent high-precision data on electroweak bosons~\cite{Khachatryan:2015hha,Alice:2016wka,Khachatryan:2015pzs,Aad:2015gta} and dijets~\cite{Chatrchyan:2014hqa,Sirunyan:2018qel} in $p$+Pb collisions, as well as charmonium production in ultra-peripheral collisions (UPCs)~\cite{Abelev:2012ba,Khachatryan:2016qhq} where the two incoming nuclei have no strong-interaction overlap. The improved constraints on nuclear PDFs from these reactions will aid in the interpretation of $D$-meson spectra in $p$+Pb collisions, and also have an impact on our understanding of HF spectra at low \pt measured in URHICs as discussed in Sec.~\ref{sec:diff}. 
 
QCD factorization has been found to work well when comparing calculations to high-\pt light-flavor hadron and HF meson, as well as electroweak-boson and dijet data at the LHC,
showing no significant modifications due to hadronization or final-state interactions in minimum-bias $p/d$+A collisions within current experimental uncertainties. On the other hand, enhancements in the electron $R_{d\rm A}$ at intermediate \pt were observed at RHIC~\cite{Adare:2013lkk}, posing a challenge to simultaneously describe the single-lepton $R_{d\rm A}$ and $D$-meson data at RHIC. More precise measurements of charm and bottom hadron production in $p/d$+A collisions at both RHIC and the LHC are necessary to disentangle initial- and final-state effects (including hadronization) in \RAA's at low \pt. 


\begin{figure}[htb]
\begin{minipage}[t]{1\linewidth}
\centering
\includegraphics[width=3.5in]{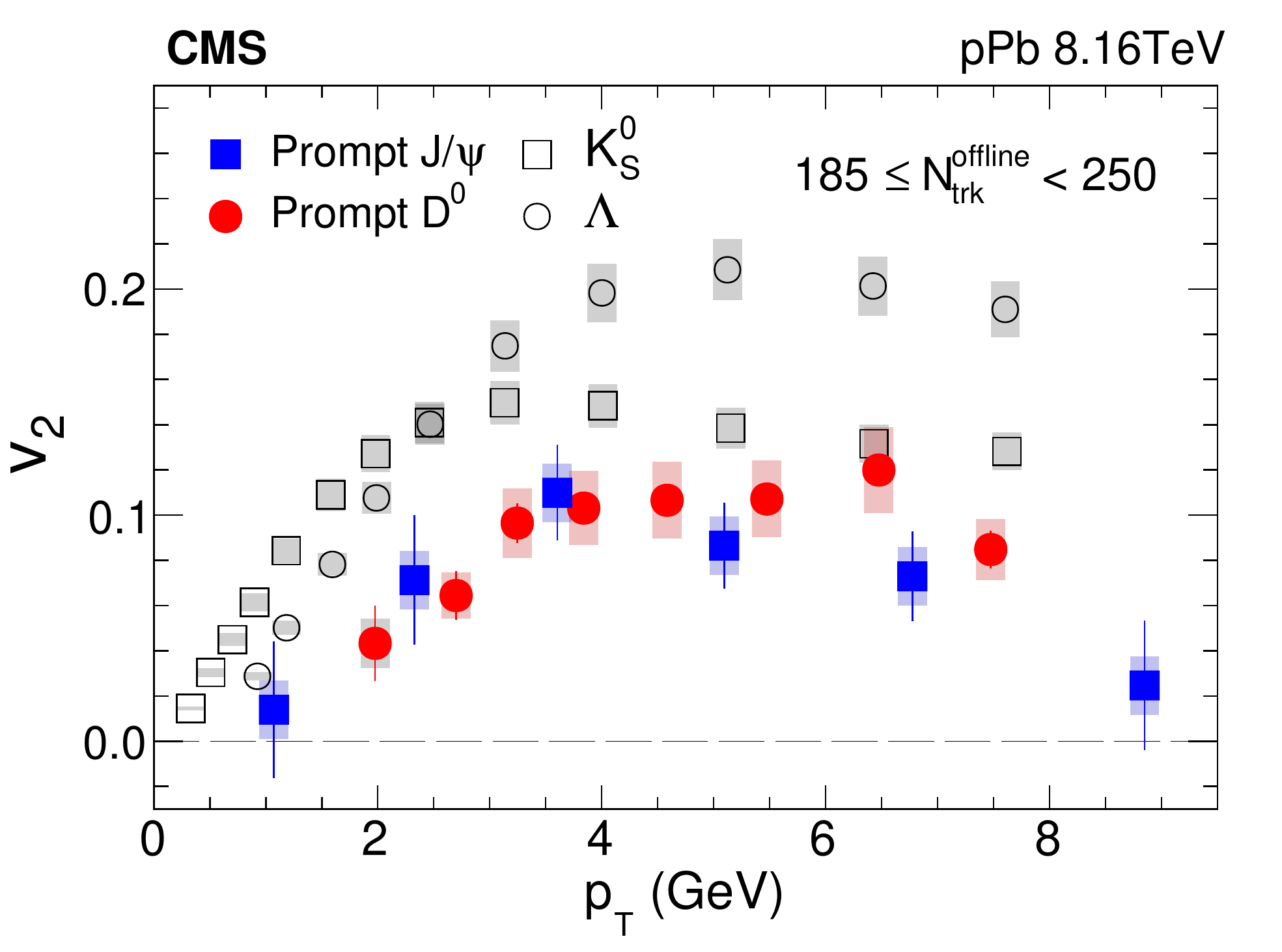}
\end{minipage}
\hspace{0.1cm}
\caption{Elliptic flow of prompt $J/\psi$ mesons (blue squares) at forward rapidities ($-2.86<y_{\rm cm}<-1.86$ and $0.94<y_{\rm cm}<1.94$), as a function of \pt\ in high-multiplicity ($185 \leq N^{\rm offline}_{\rm trk}< 250$) $p$+Pb collisions at $\sqrtsnn$\,=\,8.16\,TeV. Data for $K^0_S$ (open squares), $\Lambda$ (open circles) and prompt $D^0$ mesons (red circles) at midrapidity ($-1.46<y_{\rm cm}<0.54$) from previous CMS measurements are also shown for comparison. The error bars correspond to statistical uncertainties, while the shaded areas denote the systematic uncertainties.}
\label{fig:pACompliation}
\end{figure}

\subsection{Challenges from high-multiplicity events in small-collision systems}
\label{ssec:pA-challenge}

The studies of CNM effects with $p$+$p$ and $p/d$+A collisions usually assume that a QGP is not formed in these systems, \ie, that the hot-medium effects are negligible. However, while this may be a good approximation in MB events, it has been  challenged by signatures of collective flow (both radial and elliptic) observed in light-flavor hadron spectra in high-multiplicity $p$+$p$~\cite{Khachatryan:2010gv}, $p$+Pb~\cite{Abelev:2012ola,CMS:2012qk,Aad:2012gla,Aaij:2015qcq} and $d$+Au~\cite{Adare:2013piz,Adamczyk:2015xjc} collisions.
Yet, QGP is not the only explanation of these phenomena in small-collision systems. Initial-state parton correlations and multi-parton interactions rooted in the complex hadronic structure complicate the interpretation of $p$+$p$ and $p$+A particle-correlation measurements. It has also been suggested that surface emission, with otherwise few rescatterings in the small collision systems, could generate a sizable $v_2$ without QGP formation~\cite{He:2015hfa}. On the other hand, this escape mechanism was found to be less effective for charm quarks, which are therefore a better gauge of collectivity. Finally, an azimuthal asymmetry could be generated during initial scatterings of nuclear penetration, referred to as initial-state correlation model~\cite{Dusling:2013qoz}, where correlations originate from color flux tubes of strong fields expanding in the longitudinal direction between two outgoing ions. 

Contrary to low-\pt spectra, the measurements of charged-particle $R_{d\rm A}$'s in 0--20\% $d$+Au collisions at \sqrtsnn\,=\,200\,GeV by PHOBOS~\cite{Back:2003ns} and in 0--5\% $p$+Pb collisions at \sqrtsnn\,=\,5.02\,TeV by ALICE~\cite{Adam:2014qja} did not find significant indications of any suppression, suggesting little to no final-state interactions. This was corroborated by a recent measurement of various $D$-meson $R_{p\rm A}$'s  in central $p$+Pb collisions at \sqrtsnn\,=\,5.02\,TeV by ALICE~\cite{ALICE-PUBLIC-2017-008}.

On the other hand, recent $v_2$ measurements of $J/\psi$~\cite{Acharya:2017tfn,Sirunyan:2018kiz}, charm mesons~\cite{Sirunyan:2018toe} and HF leptons~\cite{Acharya:2018dxy} in $p$+Pb collision were rather surprising. Substantial azimuthal-anisotropy signals have been observed in high multiplicity $p$+A events, almost as large as in the light-flavor sector, cf.~Fig~\ref{fig:pACompliation}.  
The physical origin of the large flow signal in the HF flavor sector is not yet understood. To gain more insights about the underlying mechanisms, it is crucial to study possible energy loss or recombination effects on charm quarks by measuring charm-hadron spectra in ultra-central $p$+$p$, $p$+A and $d$+A collisions. The momentum balance between isolated photons and charm mesons might reveal additional information on the presence of final-state interactions. Furthermore, the angular distributions of photon-tagged relative to jet-tagged charm mesons could provide information about the origin of the large $v_2$ signal, by comparing the direction of the outgoing charm meson relative to an unmodified reference direction in $p$+$p$ (photon-tagged) or to the direction of the dominant energy flow in the tagged jet.

Finally, intermediate-size systems, such as O+O, Ar+Ar or $^3$He+A collisions, are of great interest to elaborate the transition of HF observables from small to large systems, by varying the strength of the collective flow in the bulk medium and path lengths of high energy heavy quarks~\cite{Citron:2018lsq}. 
Those collisions could be delivered in future runs at the LHC and RHIC, to search for the onset of jet quenching phenomena (thus far elusive in $p$+A collisions) while scrutinizing the development of the HF $v_2$ relative to the bulk medium.




\begin{figure}[th]
\centering
\begin{minipage}[t]{0.45\linewidth}
\includegraphics[width=1\textwidth]{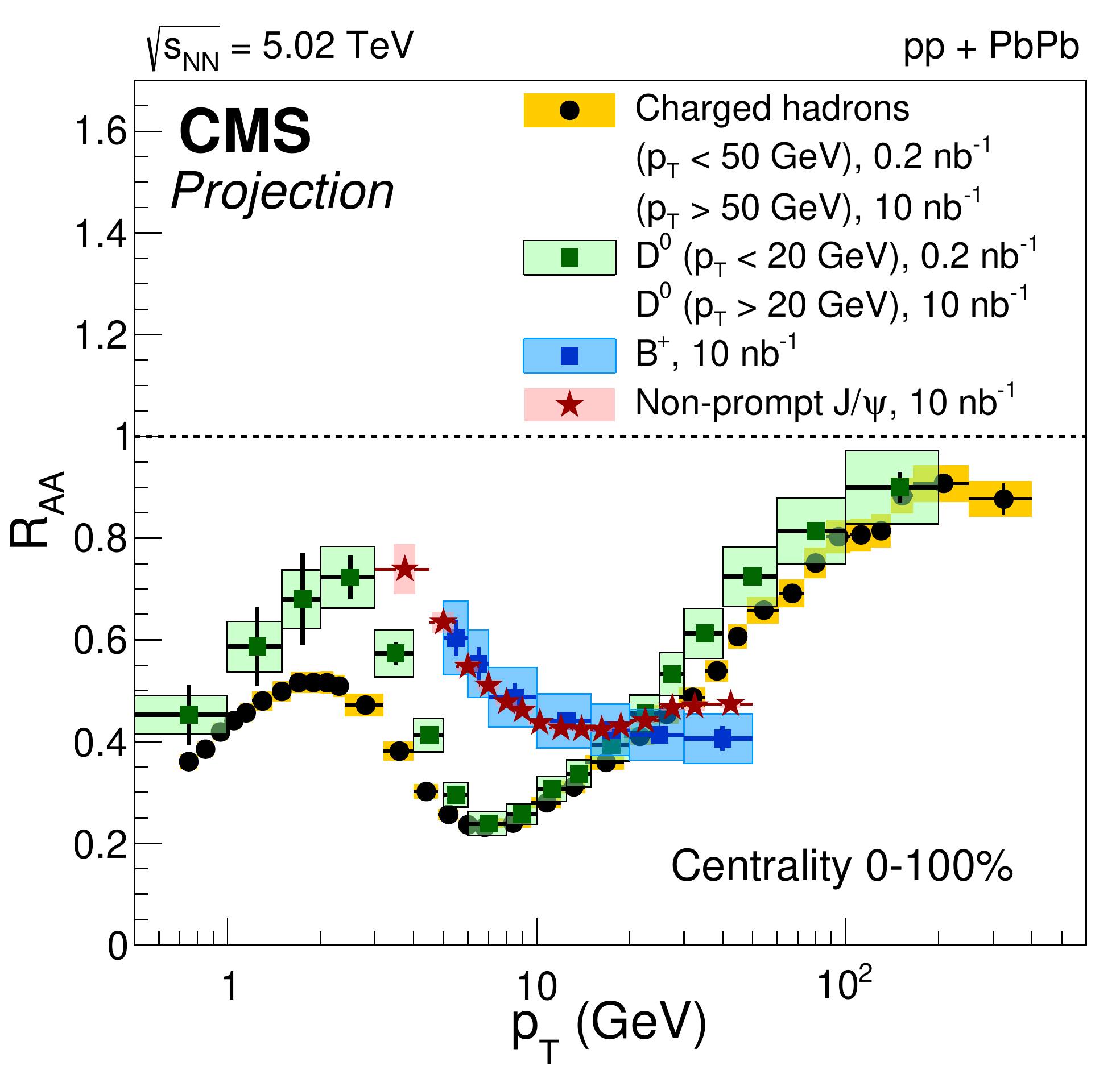}
\end{minipage}
\begin{minipage}[t]{0.5\linewidth}
\includegraphics[width=1\textwidth]{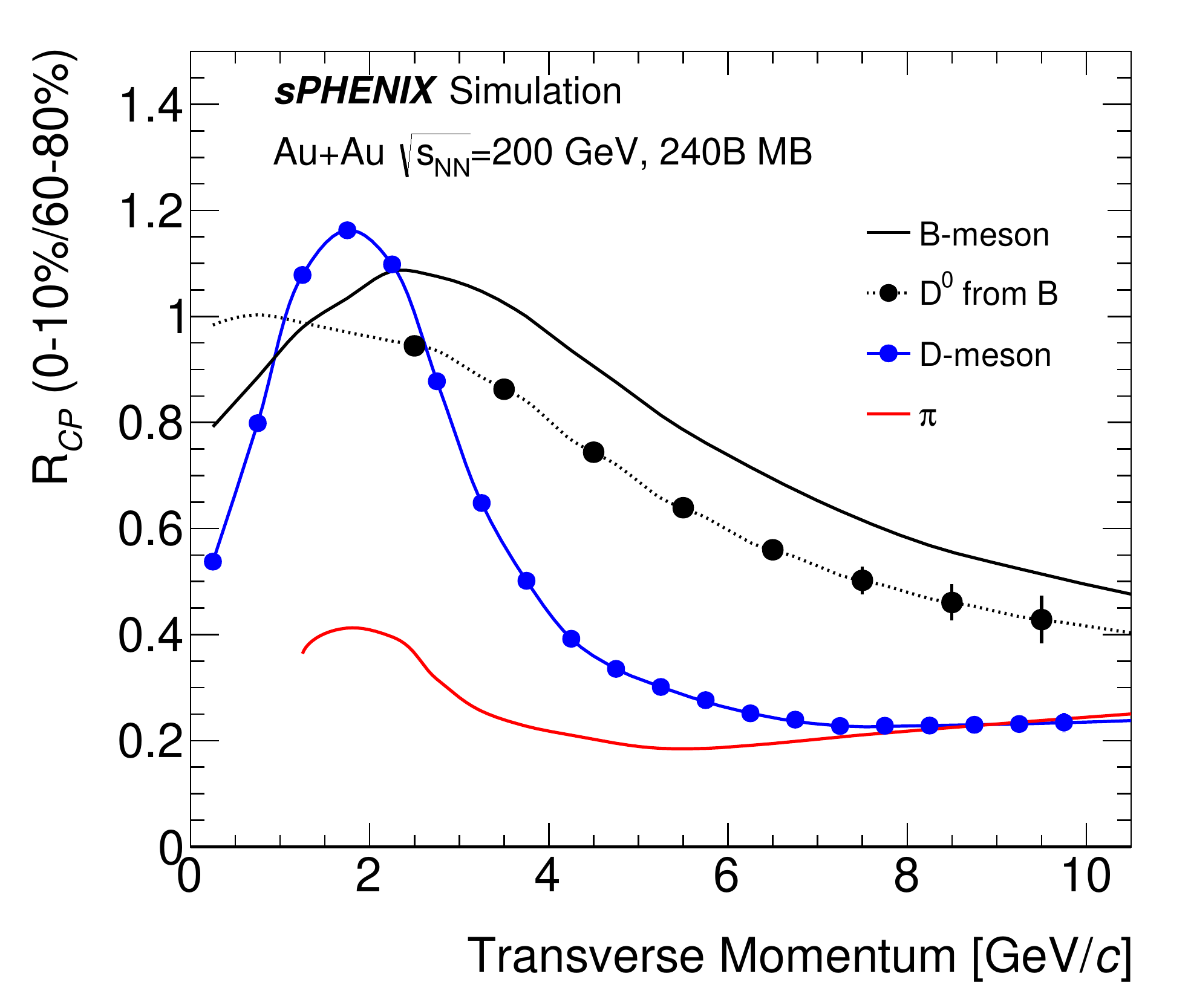}
\end{minipage}
\caption{(Left) Projected statistical and systematic uncertainties of future CMS $R_{\rm AA}$ measurements of $B^+$ hadrons, non-prompt $J/\psi$, prompt $D^0$ and inclusive charged hadrons in 0--100\% minimum bias Pb+Pb collisions at \sqrtsnn\,=\,5.02\,TeV~\cite{CMSProj}. (Right) Projected statistical uncertainties of future sPHENIX measurements of prompt and non-prompt $D^0$ $R_{\rm CP}$ (central-to-peripheral ratio) in 240 billion minimum bias Au+Au collisions at \sqrtsnn\,=\,200\,GeV~\cite{sPHENIXMVTX}. The datasets are expected to be collected in five years of RHIC physics runs between 2023-2027.}
\label{fig:ProjectedRAA}
\end{figure}

\vspace{5cm}




\section{Conclusions}
\label{sec:concl}
\begin{summary}[SUMMARY POINTS]
\begin{enumerate}
\item
Heavy-flavor particles , as a ``conserved" probe of the medium created in heavy-ion collisions, reveal fundamental information on the transport properties and the underlying microscopic interactions of QCD matter, from low-momentum diffusion to high-momentum energy loss.  
\item 
Experimental measurements of the nuclear modification factor and elliptic flow of $D$ mesons at RHIC and the LHC have revealed a remarkably strong coupling of low-momentum charm to the collectively expanding fireball medium. Transport theoretical analyses of the data have constrained the heavy-quark diffusion coefficient in the QGP near $T_{\rm pc}$ at around 2--4 times its quantum lower bound, implying scattering rates in excess of 2--3/(fm/$c$) (or widths of 0.5\,GeV or more).
\item
The $D$-meson \RAA and $v_2$ data have also revealed a transition from a diffusion dominated regime of elastic interactions to an energy loss regime at high momentum over the range of $\pt\simeq$ 5--10\,GeV/$c$. Beyond this range, the data suggest a degeneracy of $D$-meson and light-flavor hadrons, corroborating the dominance of a flavor-independent energy loss mechanism. Bottom \RAA data appear to merge into the degeneracy at $\pt\simeq$~20--30\,GeV/$c$, consistent with expectations based on the $b$- to $c$-quark mass ratio.  
\item 
The hadro-chemistry of heavy-flavor hadrons in heavy-ion collisions enables unique insights into the mechanisms of hadronization of the QGP. Systematic measurements have commenced with first results on $D_s^+/D^0$, $B_s^0/B^+$ and $\Lambda_c/D^0$ ratios, suggesting a critical role of recombination processes of heavy quarks with thermal partons from the QGP. 
\item
Heavy-flavor probes can help to scrutinize the possible formation of a fireball in small nuclear-collision system. Current data, showing a large $v_2$ but small deviations of the \RAA from one, give seemingly conflicting indications. 
\end{enumerate}
\end{summary}

\begin{issues}[FUTURE ISSUES]
\begin{enumerate}
\item 
What is the precise temperature dependence of the heavy-flavor diffusion coefficient in QCD matter? Is it universal between charm and bottom, and what does it reveal about the properties of the strongly-coupled QGP, including relations to other transport coefficients (such as viscosity and electric conductivity)?
\item 
How do the non-perturbative interactions driving the collisional diffusion of heavy quarks at low momentum give way to a perturbative regime of energy loss via gluon radiation? Can measurements of angular correlations between heavy-flavor particles disentangle the two regimes?  
\item 
Can recombination processes give a controlled description of the hadro-chemistry of heavy-flavor hadrons as a function of momentum, to understand the observed enhancement in $D_s^+$ and $\Lambda_c^+$ hadrons in heavy-ion collisions, and predict further states such as bottom baryons or the $X(3872)$ to be measured in the future? Can recombination processes offer insights into manifestations of color confinement? 
\item Can heavy-flavor probes help to determine whether a Quark-Gluon Plasma is produced in high-multiplicity events of small nuclear-collision systems?  
\item 
Upcoming detector and luminosity upgrades at RHIC and the LHC will enable precision measurements of open-bottom production, heavy-flavor baryons, as well as heavy-flavor triggered correlations, designed to give profound answers to outstanding questions in QCD matter research.
\end{enumerate}
\end{issues}

\section*{DISCLOSURE STATEMENT}
The authors are not aware of any affiliations, memberships, funding, or financial holdings that
might be perceived as affecting the objectivity of this review. 

\section*{ACKNOWLEDGMENTS}
This work has been supported by the Department of Energy, Office of Science, under grant no.~KB0201022 (XD), the Department of Energy under Early Career Award no.~DE-SC0013905 (YL), and the National Science Foundation under grant no.~PHY-1614484 (RR).

%




\bibliographystyle{ar-style5}

\bibliography{OHFHIC}
 
\end{document}